\documentclass[floatfix,prc]{revtex4}

\usepackage{dcolumn}
\usepackage{bm}
\usepackage[dvips]{color}
\usepackage{graphicx}
\usepackage{amsmath}
\usepackage{amssymb}

\newcommand{\m}{\mathbf}
\def\beq{\begin{equation}}
\def\eeq{\end{equation}}
\def\be{\begin{eqnarray}}
\def\ee{\end{eqnarray}}

\begin{document}
\title{Nuclear Effects in Neutrino Interactions and their Impact \\ on the Determination of
Oscillation Parameters}

\author{Omar Benhar and Noemi Rocco}
\affiliation
{INFN and Department of Physics, ``Sapienza'' Universit\`a di Roma \\
Piazzale Aldo Moro, 2. I-00185 Roma, Italy}

\date{\today}

\begin{abstract}

The quantitative description of the effects of nuclear dynamics on the measured neutrino-nucleus cross sections
-- needed to reduce the systematic uncertainty of long baseline neutrino oscillation experiments -- involves severe difficulties.
Owing to the uncertainty on the incoming neutrino energy, different reaction mechanisms contribute to the 
cross section measured at fixed energy and scattering angle of the outgoing lepton, and must therefore be consistently taken into account 
within a unified model. We review the theoretical 
approach based on the impulse approximation and the use of realistic nucleon spectral functions, allowing one to 
describe a variety of reaction mechanisms active in the broad kinematical range covered by neutrino experiments.
The extension of this scheme to include more complex mechanisms involving the two-nucleon current, which are
believed to be important, is also outlined. The impact of nuclear effects on the determination of neutrino oscillation 
parameters is illustrated by analyzing the problem of neutrino energy reconstruction.

\end{abstract}



\maketitle

\section{Introduction}

Experimental searches of neutrino oscillations exploit neutrino-nucleus interactions
to infer the properties of the beam particles, which are largely unknown.
The use of nuclear targets as detectors,
 while allowing for a substantial increase of the event rate,
entails non trivial problems, as the interpretation of the observed signal requires a quantitative
understanding of neutrino-nucleus interactions.
Given the present experimental accuracy,
the treatment of nuclear effects is in fact regarded as one of the
main sources of systematic uncertainty (see, e.g., Ref.\cite{NUINT11}).

\begin{figure}[h!]
\includegraphics[scale= 0.465]{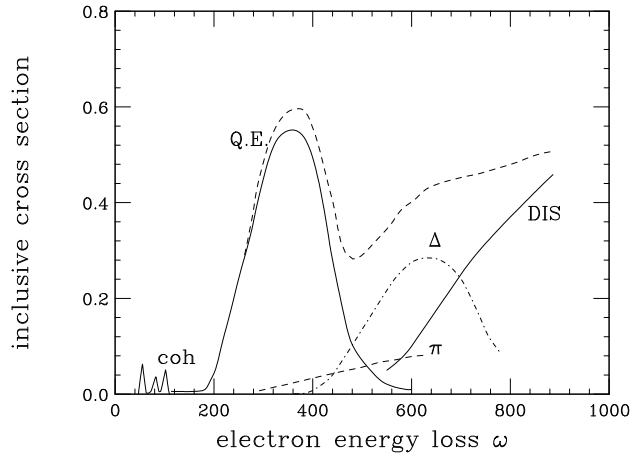}
\vspace*{-.1in}
\caption{Schematic representation of the inclusive electron-nucleus cross section at beam energy around 1 GeV, as a function of energy loss.}
\label{regime}
\end{figure}

Over the past decade, a growing effort has been made, aimed at making use of the knowledge of the nuclear response acquired  from 
experimental and theoretical studies of electron scattering.
Electron-nucleus scattering cross sections are usually analyzed at fixed beam energy $E_e$, and electron scattering angle $\theta_e$ as a function of the energy loss $\omega$.
 As an example, Fig.\ref{regime} shows the typical behavior of the double differential cross section of the inclusive process
 \beq
 \label{firsteq}
 e + A \rightarrow e^\prime + X \ ,
 \eeq
 in which only the outgoing lepton is detected,  at beam energy around 1 GeV. Here, $A$ and $X$ denote the target nucleus, in its ground state, and the undetected 
 hadronic final state, respectively.

 It is apparent that the different reaction mechanisms, yielding the dominant contributions to the cross section at different values of $\omega$ (corresponding to
 different values of the Bjorken scaling variable $ x=Q^2/2M\omega$, where $M$ is the nucleon mass and
 $Q^2=4E_e(E_e-\omega) \sin^2\theta_e/2$) can be easily identified.

The bump centered at $\omega\sim Q^2/2M$, or $x\sim1$, the position and width of which are determined by the momentum and removal energy distribution of the struck particle, corresponds to single nucleon knockout, while the structure visible at larger $\omega$ reflects the onset of coupling to two-nucleon currents, arising from meson exchange processes, excitation of nucleon resonances and deep inelastic scattering.

The available theoretical models of electron-nucleus scattering provide an overall satisfactory description of the data over a broad kinematical range.
In particular, in the region in which quasi elastic scattering dominates, the data is generally reproduced with an accuracy of few percent (for a recent review on
electron-nucleus scattering in the quasi elastic sector, see Ref. \cite{RMP}).

\begin{figure}[h!]
\includegraphics[scale= 0.5]{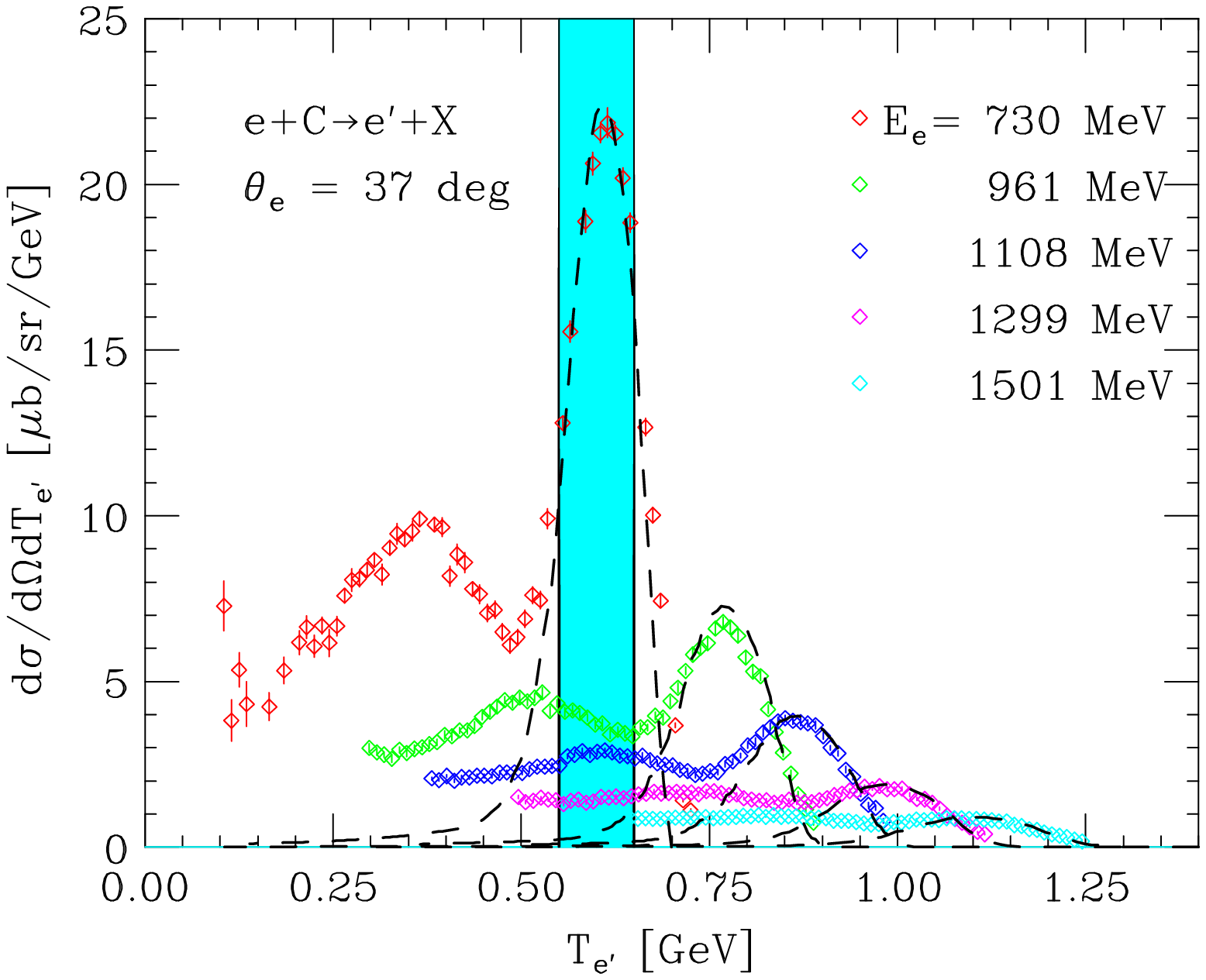} \hspace*{.5in} \includegraphics[scale= 0.475]{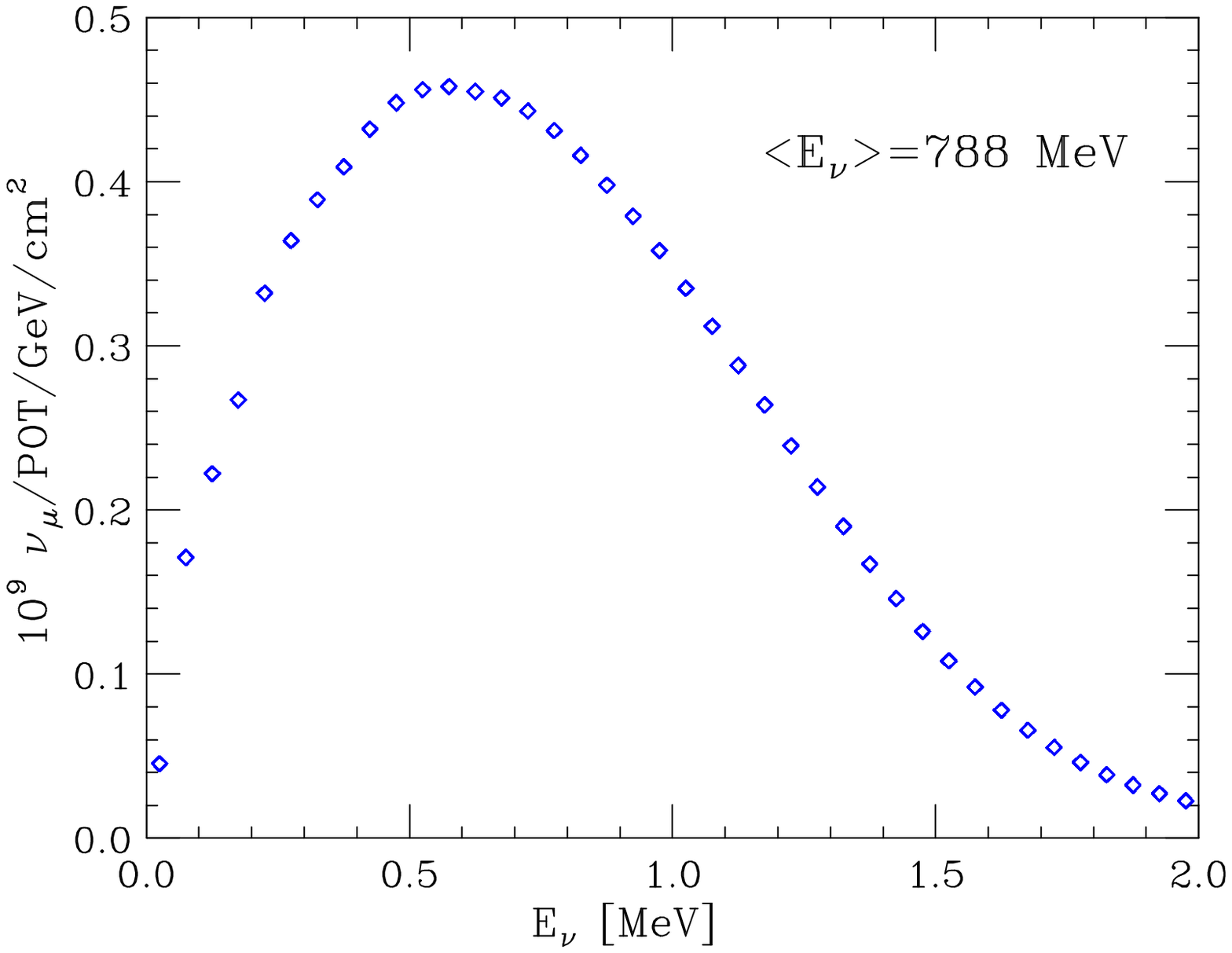} 
\vspace*{-.1in}
\caption{Left panel: inclusive electron-carbon cross sections at $\theta_e = 37$ deg and beam energies
ranging between 0.730 and 1.501 GeV \cite{12C_ee,12C_ee_2}, plotted as a function of the energy of the outgoing electron. Right panel:
 energy dependence of the MiniBooNE neutrino flux \cite{BooNECCQE}.}
\label{fluxav}
\end{figure}
  
Because neutrino beams are always produced as secondary decay products, their energy is not sharply defined, but broadly distributed.
As a consequence, in charged-current neutrino scattering processes detecting the energy of the outgoing lepton, $T_\ell$,  {\em does not} provide a 
measurement of the energy transfer, $\omega$, and different reaction mechanisms can contribute to the double differential cross section at 
fixed $T_\ell$ and lepton scattering angle, $\theta_\ell$. This feature is illustrated in the left panel of Fig. \ref{regime}, showing the inclusive 
electron-carbon cross sections at  $\theta_e = 37$ deg and beam energies ranging between 0.730 and 1.501 GeV, as a function of energy of the 
outgoing electron \cite{12C_ee,12C_ee_2}. It clearly appears that the highlighted  $550 < T_{e^\prime} < 650$ MeV bin, corresponding to quasifree
kinematics at $E_e = 730 \ {\rm MeV}$, picks up contributions from scattering processes taking place at different beam energies, in which 
reaction mechanisms other than single nucleon knockout are known to be dominant. To gauge the extent to which different 
contributions are mixed up in a typical neutrino experiment, consider the energy distribution of the MiniBooNE neutrino flux, displayed 
in the right panel of  Fig. \ref{fluxav}, showing that the fluxes corresponding to energies $E_\nu =$ 730 and and  961 MeV are within less than 
20\% of one another. It follows that, if we were to average the electron-carbon data of the left panel with the flux of the right panel, the 
cross sections corresponding to beam energies 730 and 961 MeV would contribute to the measured cross section in the highlighted bin with about 
the same weight. 

The above discussion implies that the understanding of the flux averaged neutrino cross section requires the development of 
theoretical models providing a consistent treatment of all reaction mechanisms active in the broad kinematical range 
corresponding to the relevant neutrino energies.

In Section \ref{nuA:xsection} we discuss the structure of the neutrino-nucleus cross section, and point out that a consistent 
treatment of relativistic effects and nucleon-nucleon correlations 
requires the factorization of the nuclear vertex. The main elements of the resulting expression of the cross section, i.e. 
the nucleon spectral function and the elementary neutrino-nucleon cross section, are also analyzed. In Section \ref{nsf} we briefly 
review the available empirical information on the nucleon weak structure functions in the kinematical regimes corresponding 
to quasi elastic scattering, resonance production and deep inelastic scattering, while Section \ref{interpretation} is devoted to 
a discussion of the ambiguities implied in the interpretation of the events labeled as quasi elastic. As an example of the impact
of nuclear effects on the determination of neutrino oscillations, in Section \ref{reconstruction} we analyze the problem of neutrino 
energy reconstruction. Finally, in Section~\ref{conclusions} we summarize the main issues and state our conclusions. 

\section{The neutrino-nucleus cross section}
\label{nuA:xsection} 

Let us consider, for definiteness, charged-current neutrino-nucleus interactions. The formalism discussed in this section can be readily generalized 
to the case of neutral current interactions \cite{veneziano}. The double differential cross section of the 
process (compare to Eq. \eqref{firsteq})
\beq
\nu_\ell + A \to \ell^- + X \ ,
\eeq
can be written in the form \cite{NPA}
\begin{align}
\label{nucl:xsec}
\frac{d^2\sigma}{d\Omega_{{\bf k}^\prime} dk_0^\prime}& =\frac{G_F^2\,V^2_{ud}}{16\,\pi^2}\,
\frac{|\bf k^{\prime}|}{|\bf k|}\,L_{\mu\nu}\, W_A^{\mu\nu} \ .
\end{align} 
In the above equation, $k \equiv (k_0,{\bf k})$ and $k^\prime \equiv (k^\prime_0,{\bf k}^\prime)$
are the four momenta carried by the incoming neutrino and the outgoing charged lepton, respectively, 
$G_F$ is the Fermi constant and $V_{ud}$ is the CKM matrix element coupling $u$
and $d$ quarks. The tensor $L_{\mu\nu}$, defined as (we neglect the term proportional to $m_\ell^2$, 
where $m_\ell$ is the mass of the charged lepton)
\begin{align}
\label{leptensor}
L_{\mu\nu}&=8\,\left[k_\mu^{\prime}\,k_\nu+k_\nu^{\prime}\,k_\mu- g_{\mu\nu}(k\cdot
k^{\prime})-i\,\varepsilon_{\mu\nu\alpha\beta}\,k^{\prime \beta}\,k^\alpha \right] \ , 
\end{align}
is completely determined by the lepton kinematics, whereas the nuclear tensor
$W_A^{\mu\nu}$, containing all the information on strong interaction dynamics,
describes the response of the target nucleus. Its definition
\begin{align}
\label{hadronictensor}
W_A^{\mu\nu}&= \sum_X \,\langle 0 | {J_A^\mu}^\dagger | X \rangle \,
      \langle X | J_A^\nu | 0 \rangle \;\delta^{(4)}(p_0 + q - p_X) \ ,
\end{align}
with $q=k-k^\prime$, involves the target initial and final states $|0\rangle$ and $|X\rangle$, carrying four momenta $p_0$ and $p_X$, 
respectively, as well as the nuclear  current operator 
\beq
\label{nuclear:current}
J_A^\mu= \sum_i j^\mu_i+\sum_{j>i} j^\mu_{ij} \ ,
\eeq
where $j^\mu_{ij}$ denotes the two-nucleon contribution arising from meson-exchange processes. 

In the kinematical region corresponding to low momentum transfer, typically $|{\bf q}| < 400 \ {\rm MeV}$, in which non relativistic approximations 
are expected to work, the tensor of Eq. \eqref{hadronictensor} can be evaluated within highly realistic nuclear models \cite{schiavilla,lovato12C}. 
However, the event analysis of accelerator-based neutrino experiments requires theoretical approaches that can be applied in the relativistic regime. 
The importance of relativistic effects can be easily grasped considering that
the mean momentum transfer of quasi elastic (QE) processes obtained by averaging over the MiniBooNE \cite{BooNECCQE} and Miner$\nu$a \cite{Minerva} neutrino fluxes  
turn out to be $\sim 640$ and $\sim 880$ MeV, respectively.

Non relativistic nuclear many-body theory, based on dynamical models
strongly constrained by phenomenology, provides a fully consistent theoretical approach allowing for an accurate description of the target initial  state, independent of momentum transfer. 
On the other hand,  at large $|{\bf q}|$ the treatment of both the nuclear current and the hadronic final state unavoidably requires
approximations.  

\subsection{The impulse approximation}
\label{IA}
 The Impulse Approximation (IA) scheme, extensively employed to analyze electron-nucleus scattering data \cite{RMP}, is based on the tenet that, 
at momentum transfer $\m{q}$ such that $\m{q}^{-1} << d$, $d$ being the average nucleon-nucleon distance in the target,  
neutrino-nucleus scattering reduces to the incoherent sum of scattering processes involving individual nucleons. Moreover, final state interactions between the outgoing hadrons and the spectator nucleons are assumed to be negligible. 

Within the IA picture, the nuclear current of Eq.\eqref{nuclear:current} reduces to the sum of one-body terms,  
while the final state simplifies to the direct product of the hadronic state produced at the interaction vertex, with momentum ${\bf p}_x$, and the state
describing the $(A-1)$-nucleon residual system, carrying momentum ${\bf p}_R$, i.e.
\beq
\label{factorization}
|X\rangle \longrightarrow |x,{\bf p}_x \rangle \otimes |R, {\bf p}_R\rangle \ ,
\eeq
implying 
\beq
\sum_X |X\rangle\langle X|\rightarrow \sum_x\int \,{d^3p_x}|x, \m{p}_x\rangle\langle \m{p}_x,x|  \sum_R   \ \int d^3p_R|R,\m{p}_R\rangle\langle \m{p}_R,R| \ .
\eeq
Insertion of a complete set of free nucleon states, satisfying
\beq
\int \,{d^3k}|N,\m{k}\rangle\langle\m{k},N|= 1,
\eeq
leads to the factorization of the nuclear current matrix element according to
\beq
\label{fact:matel}
\langle 0|J^{\mu}_A|X\rangle=  \Big( \frac{M}{\sqrt{|\m{p}_R|^2 + M^2}}\Big)^{1/2} \langle 0|R,\m{p}_R;N,-\m{p}_R\rangle
  \sum_i \langle -\m{p}_R,N|j^{\mu}_i|x,\m{p}_x\rangle  \ ,
\eeq
where the factor $(M / \sqrt{|\m{p}_R|^2 + M^2})^{1/2}$, with $M$ being the nucleon mass, takes into account the implicit covariant normalization of the state $\langle -\m{p}_R, N|$ in the matrix element of $j^{\mu}_i$.

Using the above relations, the hadronic tensor can be rewritten in the form
\beq
\label{tensorlong}
\begin{split}
W^{\mu\nu} & = \sum_{R,x} \int \,{d^3p_Rd^3p_x}|\langle0|R,\m{p}_R;N,-\m{p}_R\rangle|^2 
 \left( \frac{M}{\sqrt{|\m{p}_R|^2 + M^2}}\right) \\ 
& \times  \sum_i \langle -\m{p}_R, N|{j^{\mu}}^\dagger_i|x,\m{p}_x\rangle\langle \m{p}_x,x|j^{\nu}_i|N,-\m{p}_R\rangle  
 \delta^3(\m{q} - \m{p}_R - \m{p}_x)\delta(\omega + E_0 - E_R - E_x) \ ,
\end{split}
\eeq
where $E_0$ is the target ground state energy, $E_R= \sqrt{|\m{p}_R|^2 + M_R^2}$, $M_R$ being the mass of the recoiling system, and $E_x$ is the energy of the hadronic state $x$
produced at the interaction vertex.

Equation \eqref{tensorlong} can be cast in the more concise form
\beq
W^{\mu\nu}_A =   \int \,{d^3p \ dE} \ P(E, \m{p}) 
\frac{M}{E_p} \sum_x \langle \m{p}, N|j^{\mu}_i | x, \m{p} + \m{q}\rangle\langle x, \m{p} + \m{q}|j^{\nu}_i |\m{p}, N\rangle  \delta(\tilde{\omega} + \sqrt{\m{p}^2 + M^2} - E_x) \ ,
\eeq
where 
\beq
\tilde{\omega}= E_x - \sqrt{\m{p}^2 + M^2} = \omega + M - E - \sqrt{\m{p}^2 + M^2}= \omega + E_0 - E_R - \sqrt{\m{p}^2 + M^2} \ ,
\eeq
and the spectral function
\beq
\label{spectfunc}
P(\m{p},E)= \sum_R |\langle 0|R, -\m{p}\rangle|^2\delta(E - M + E_0 - E_R) \ ,
\eeq
yields the probability of removing a nucleon with momentum $\m{p}$ from the target ground state leaving the residual system with excitation energy $E$.

Using the  definition of the tensor describing the interactions of the \emph{i}-th nucleon in free space (the subscripts $\alpha = p, n$ denote
proton and neutron, respectively)
\beq
\label{N:tens}
\mathcal{W}_\alpha^{\mu\nu} =  \sum_x \langle \m{p}, N|{j^{\mu}}^\dagger_\alpha |X, \m{p}+\m{q} \rangle\langle x,\m{p}+\m{q}|j^{\nu}_\alpha |\m{p}, N\rangle
\delta(\tilde{\omega} + \sqrt{\m{p}^2 + M^2} - E_x) \ ,
\eeq
we finally obtain, for a target nucleus with $Z=A/2$ 
\beq 
W^{\mu\nu}_A= A  \int \,{d^3p \ dE}  \  \frac{M}{E_p } \ P(\m{p},E) \  \mathcal{W}_N^{\mu\nu}
\eeq
with $ \mathcal{W}_N^{\mu\nu} = ( \mathcal{W}_p^{\mu\nu} + \mathcal{W}_n^{\mu\nu} )/2$. 
Note that in deriving the above equation we have made the assumption, largely justified in isoscalar nuclei, that the proton and neutron 
spectral functions be the same.

It has to be emphasized that the replacement of $\omega$  with $\tilde{\omega}$ (see Eq.\eqref{N:tens}) is meant to take into account the fact that a fraction $\delta\omega$ of the 
energy transfer goes into excitation energy of the spectator system. Therefore, the elementary scattering process can be described as if it took place in free space with energy transfer 
$\tilde{\omega}= \omega-\delta\omega$. 

Collecting the above results, the nuclear cross section can be finally written in the transparent form
\begin{equation}
\label{sigma1}
\frac{d^2\sigma_{IA}}{d\Omega_{{\bf k}^\prime} d k_0^\prime}= A \ \int \,d^3p \ dE \ P(p,E) \frac{d^2\sigma_{elem}}{d\Omega_{{\bf k}^\prime}dk_0^\prime}  ,	
\end{equation}
with
\begin{equation}
\label{sigma2}
\frac{d^2\sigma_{elem}}{d\Omega_{{\bf k}^\prime}dk_0^\prime}= \frac{G^2_F V^2_{ud}}{16\pi^2} \frac{|\m{k^\prime}|}{|\m{k}|}\frac{1}{4E_{|\m{p}|}E_{|\m{p+q}|}} L_{\mu \nu}
\mathcal{W}_N^{\mu \nu} \ .
\end{equation}

In conclusion, within the IA scheme it is possible to trace back the hadronic tensor corresponding to the nuclear target to the ones describing the elementary 
interaction with isolated nucleons -- which can be, at least in principle, measured using proton and deuteron targets -- provided the four momentum transfer $q$ is replaced 
with ${\tilde q}\equiv({\tilde \omega},{\bf q})$ and an integration on the nucleon momentum and removal energy is carried out,  with a weight given by the spectral function.

We emphasize that, as will be discussed below, Eqs. \eqref{sigma1} and \eqref{sigma2} 
can be applied to a variety of reaction mechanisms,  including QE scattering, resonance production and Deep Inelastic Scattering (DIS).

\subsection{Spectral function}

The calculation of the target spectral function requires a model of nuclear dynamics.
The simulation codes employed for the analysis of neutrino oscillation experiments are largely based on the relativistic Fermi gas model (RFGM) \cite{Moniz3},
according to which the target nucleus can be described as a degenerate gas of protons and neutrons obeying on-shell relativistic kinematics. Nucleons occupy  
all states with momenta smaller than the Fermi momentum $p_F$ -- belonging to the Fermi sea -- and are bound with constant energy $\epsilon$. The values of these two parameters 
are determined through a fit of the position and width of the quasi elastic peak of the measured electron-nucleus scattering cross sections \cite{Moniz1,Moniz2}.

Within the RFGM, the nuclear spectral function, defined in Eq. \eqref{spectfunc}, can be written 
in the simple form
\begin{equation}
P_{RFGM}(p,E)= \frac{6\pi^2}{p^3_F}  \ \theta(p_F - p) \ \delta(E_p - \epsilon + E) \ ,
\end{equation}
where $E_p = \sqrt{ |{\bf p}|^2 + M^2 }$.

Electron scattering data have provided overwhelming evidence that the energy-momentum distribution of nucleons in the nucleus is quite different from the one predicted 
by the RFGM. The differences are to be ascribed to the presence of nucleon-nucleon (NN) correlations, mainly arising from the strongly 
repulsive nature of the NN interactions at short distances. 
Dynamical correlations give rise to virtual scattering processes leading to the excitation of the 
participating nucleons to states of energy larger than the Fermi energy, thus depleting the single particle 
levels within the Fermi sea. Owing to the contribution of nucleons belonging to a correlated pair, the 
nuclear spectral function $P({\bf p},E)$ exhibits tails extending to the region $ |p| \gg p_F$ and 
$E \gg \epsilon$.

Highly accurate theoretical calculations of the spectral function can be carried out for uniform nuclear matter, exploiting the simplifications 
arising from translation invariance  \cite{BFF}. The results 
of these calculations have 
been combined with the information obtained from coincidence $(e,e^\prime p)$ experiments  at moderate energy and momentum transfer, to obtain spectral functions of a variety of nuclei within the local 
density approximation (LDA) \cite{LDA,PRD}. 

According to the LDA scheme, the spectral function is written in the form 
\beq
P_{LDA}(\m{p}, E)= P_{MF}(\m{p}, E) + P_{\rm corr}(\m{p}, E) \ ,
\eeq
where the two terms describe the contributions associated with the nuclear mean field and NN correlations, respectively.  
The former is usually written in the factorized form
\beq
P_{MF}(\m{p}, E)= \sum_{n \in \{F\}} Z_n|\phi_n(\m{p})|^2F_n(E - E_n) \ ,
\eeq 
where the sum is extended to all states belonging to the Fermi sea, while the spectroscopic factor $Z_n < 1$ and the function $F_n(E - E_n)$, accounting for the finite width of the $n$-th shell-model state, take into account the effects of 
residual interactions not included in the mean-field picture. In the absence of all interactions,  $Z_n~\rightarrow~1$ and  $F(E-E_n)~\rightarrow~\delta(E~-~E_n)$. 

The correlation contribution is given by 
\beq
P_{\rm corr}(\m{p}, E)= \int \,{d^3r} \varrho_A(\m{r})P^{NM}_{\rm corr}(\m{p}, E;\varrho= \varrho_A(\m{r})) \ ,
\eeq
where $\varrho_A(\m{r})$ is the nuclear density distribution and $P^{NM}_{corr}(\m{p},E;\varrho)$ is the correlation part of the spectral function of  nuclear 
matter at uniform density $\varrho$. Note that the spectroscopic factors $Z_n$ are constrained by the normalization requirement 
\beq
\int \,{d^3p \ dE}\ P_{LDA}(\m{p}, E)= 1 \ .
\eeq
Typically, the mean-field contribution accounts for $\sim 80$ \% of the above normalization integral. The correlation strength located at large $p$ and $E$ has been 
recently measured at JLab using a Carbon target \cite{daniela}. The results of this analysis are consistent with the data at low missing energy and missing momentum, 
as well as with the results of theoretical calculations carried out within nuclear many-body theory.

\subsection{Neutrino-nucleon vertex}

The most general expression of the target tensor of \eqref{N:tens}
can be written in terms of five structure functions, depending on the Lorentz scalars $q^2$ and $(p \cdot q)$ only, as
\beq
\begin{split}
\label{W2}
\mathcal{W}^{\mu\nu} & = 
-g^{\mu\nu}W_1(q^2) + \frac{p^{\mu}p^{\nu}}{M^2}W_2(q^2) \\ 
& -i\epsilon^{\mu\nu\varrho\sigma}\frac{p_{\varrho}q_{\sigma}}{2M^2}W_3(q^2) + \frac{q^{\mu}q^{\nu}}{M^2}W_4(q^2) 
+ \frac{p^{\mu}q^{\nu} + p^{\nu}q^{\mu}}{M^2}W_5(q^2) \ .
\end{split}
\eeq
In scattering processes involving isolated nucleons, the structure functions $W_4$ and $W_5$ give vanishing contributions
to the cross section, after contraction of the target tensor  with $L_{\mu \nu}$. Owing to the replacement $q \to {\tilde q}$ in the 
arguments of $\mathcal{W}^{\mu \nu}$,  dictated by the IA, in neutrino-nucleus scattering this is no longer the case. However, the 
results of numerical calculations suggest that the contribution of the terms involving $W_4$ and $W_5$ is small, 
and can be safely neglected \cite{NPA}. 

The contraction of the above tensor with $L_{\mu \nu}$ of Eq. \eqref{leptensor} can be cast in the form
\beq
L^{\mu\nu}W_{\mu\nu}= 16\sum_i W_i\Big( \frac{A_i}{M^2}\Big) \ ,
\eeq
with the kinematical factors $A_i$ given by \cite{NPA}.
\begin{align}
A_1 = & M^2 \ (k \cdot k^\prime) \ \ \ , \ \ \  A_2 = (k \cdot  p)\,(k^\prime \cdot p)-\frac{A_1}{2} \ \ \ , \ \ \ 
A_3 =(k\cdot  p)\,(k^\prime \cdot  \tilde q)-(k \cdot \tilde q)\,(k^\prime \cdot  p) \\
\nonumber
A_4 & =(k\cdot \tilde q)\,(k^\prime \cdot \tilde q)-\frac{\tilde q^2}{2}\,\frac{A_1}{M^2} \ \ \ , \ \ \ 
A_5=(k\cdot  p)\,(k^\prime \cdot \tilde q)+(k^\prime \cdot  p)\,(k\cdot \tilde q)-(\tilde q \cdot p)\,\frac{A_1}{M^2} \ .
\label{def:A}
\end{align}

\section{Nucleon structure functions}
\label{nsf}

As already stated, the formalism based on the IA provides a unified framework, suitable to describe neutrino-nucleus
interaction in different kinematical regimes. In this Section, we discuss the form of the structure functions  $W_i$ in the QE and DIS regimes, 
and briefly outline the  extension of the QE form to the case of resonance production.

\subsection{Charged-Current Quasi Elastic (CCQE) scattering and resonance production}
\label{CCQE}

In the CCQE channel, the structure functions involve the energy conserving $\delta$-function enforcing the condition that  the 
scattering process be elastic. They are conveniently written in the form
\beq
\label{def:Wtilde}
W_i = {\widetilde W}_i \ \delta\Big(\tilde{\omega} + \frac{\tilde{q}^2}{2M}\Big) \ ,
\eeq
where the ${\widetilde W}_i$ can be obtained from the matrix elements of the nucleon current. Exploiting the 
CVC hypotesis and PCAC, the resulting structure functions, can be written in terms of the electromagnetic form factors, 
$F_1$ and $F_2$ and the axial form factor $F_A$ according to
\begin{align}
\label{strfunc}
{\widetilde W}_1= 2[F_A^2(1 + \tau) & + \tau(F_1 + F_2)^2] \ \ \ , \ \ \  {\widetilde W}_2= 2[F_A^2 + F_1^2 + \tau F_2^2] \ \ \ , \ \ \ {\widetilde W}_3= 2F_A(F_1 + F_2)  \ , \\ 
{\widetilde W}_4 &= [F_2^2(1 + \tau) - 2F_2(F_1 + F_2)]/2 \ \ \ , \ \ \ {\widetilde W}_5= W_2/2 \ ,
\end{align}
with $\tau = - q^2/4M^2$.

The form factors appearing in the vector current, $F_1(q^2)$ and $F_2(q^2)$, are obtained from the measured electric and magnetic nucleon form factors, $G_E$ and $G_M$, through the relations
\begin{align}
F_1(q^2)= \frac{1}{(1 -\tau)} [ G_E(q^2) - \tau G_M(q^2) ] \ \ \ , \ \ \ F_2(q^2)= \frac{1}{(1 -\tau)}[ -G_E(q^2) + G_M(q^2)] \ .
\end{align}

While more refined parametrizations of the large body of data are available (for a review, see, e.g., Ref. \cite{VFF}), the form factors $G_E$ and $G_M$ are often written in the simple dipole form
\begin{align}
G_E(q^2)= \Bigg( 1 - \frac{q^2}{M_V^2}\Bigg)^{-2} \ \ \ \ , \ \ \ \ 
G_M(q^2)= (\mu_p - \mu_n)\Bigg( 1 - \frac{q^2}{M_V^2}\Bigg)^{-2} \ ,
\end{align}
with $M_V^2= 0.71\ \rm{GeV}^2$. The axial form factor, $F_A$, is also written in the same form
\beq
\label{FA}
F_A(q^2)= g_A\Bigg( 1 - \frac{q^2}{M_A^2}\Bigg)^{-2} \ .
\eeq
The value of the axial coupling constant, $g_A= -1.261 \pm 0.004$, is obtained from neutron $\beta$-decay, while the axial mass extracted from low-statistics
neutrino-deuteron scattering data is $M_A= 1.032 \pm 0.036$ GeV \cite{bernard,bodek2}.
The contribution  of the pseudoscalar form factor, $F_P$, can be safely neglected, except for the case of $\nu_\tau$ scattering.

The generalization of the above formalism to describe the resonance production region only involves minor changes. 
Unlike the CCQE case, the structure functions depend on both $q^2$ and $W^2$, the squared invariant mass of the hadronic final state, and  the energy conserving $\delta$-function in 
Eq.\eqref{def:Wtilde} is replaced by the Breit-Wigner factor
\beq
 \frac{M_R\Gamma_R}{\pi}\frac{1}{(W^2-M_R^2)^2 + M_R^2\Gamma_R^2} \ ,
\eeq
where $M_R$ and $\Gamma_R$ denote the resonance mass and its decay width, respectively. In addition, 
the nucleon form factors are replaced by the transition matrix elements of the nucleon weak current \cite{NPA,LP}. 
 
\begin{figure}[h!]
 \includegraphics[scale= 0.5]{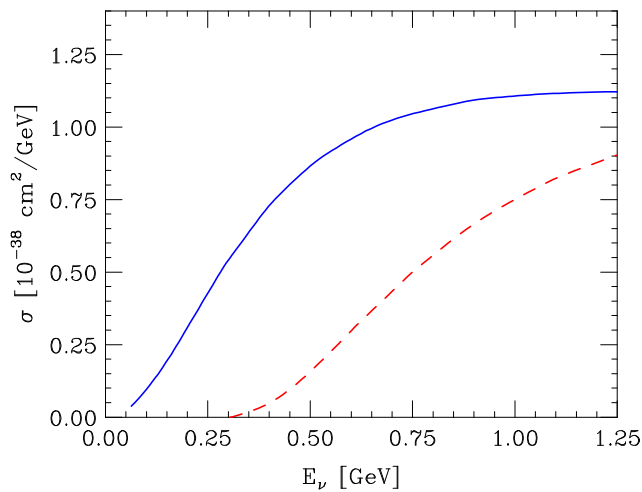} 
\vspace*{-.1in}
\caption{QE (solid line) and resonance production (dashed line) contributions to the charged-current neutrino-nucleon scattering cross
section (adapted from Ref. \cite{NPA}).}
\label{sigtot}
\end{figure}

The CCQE and resonance contributions to the total neutrino-nucleon cross section reported by the authors of Ref.~\cite{NPA} are shown 
in Fig. \ref{sigtot} as a function of neutrino energy. The resonance-production cross section has been obtained taking into account both
the $\Delta$ resonance and the three isospin $1/2$ states lying in the so-called second resonance region: the $P_{11}(1440)$,  $D_{13}(1520)$
and $S_{11}(1535)$. It clearly appears that at beam energies $\sim 1$ GeV, QE scattering and resonance production turn out to be comparable. 

The decay of the $\Delta$ resonance is a prominent mechanism leading to the appearance of pions in the final state. A detailed discussion of
 both coherent and incoherent pion production can be found in Refs.~\cite{sajjad,leitner}. 

\subsection{Deep inelastic scattering}
\label{DIS}

From the observational point of view, the DIS regime corresponds to hadronic final states with more than one pion.

In principle, the three nucleon structure functions entering  the definition of the IA nuclear cross section, Eqs. \eqref{sigma1} and \eqref{sigma2}, 
can be obtained combining neutrino and antineutrino scattering cross sections. However, as the available structure functions have been 
extracted from nuclear cross sections (see, e.g., Ref. \cite{CDHS}), 
their use in ``bottom up'' theoretical studies aimed at identifying nuclear effects involves obvious conceptual difficulties.

An alternative approach, allowing one to obtain the structure functions describing DIS on isolated nucleons, can be developed within the  
framework of the quark-parton model, exploiting the large database of DIS data collected using charged lepton beams, hydrogen and deuteron targets (see, e.g., Ref. \cite{roberts}). 
Within this scheme,  the function $F_2^{\nu N} = \omega W_2$, where $\omega$ is the energy transfer and  $W_2$ is the weak structure function of  an isoscalar nucleon, 
defined by Eq. \eqref{W2}, can be simply related to the corresponding structure function extracted from electron scattering data, $F_2^{e N}$, through
\begin{align}
F_2^{\nu N} & = \frac{18}{5} \ F_2^{e N} \ .
\end{align}
In addition, the relation 
\begin{align}
x \omega W_3 & = V(x)  \ , 
\end{align}
where $x$ is the Bjorken scaling variable and $V(x)$ denotes the valence quark distribution, implies
\begin{align}
x \omega W_3  = F_2^{eN} - 2 {\bar q}(x)  \ , 
\end{align}
${\bar q}(x)$ being the antiquark distribution. 

Using the above results and the relation $F_2 = 2xF_1$, with $F_1 = M W_1$,  one can readily obtain the weak
structure functions from the existing parametrization 
of the electromagnetic structure functions and the antiquark distribution (see, e.g., Ref. \cite{BR}).

The above procedure rests on the tenet, underlying the IA scheme, that the elementary neutrino-nucleon interaction is {\em not} affected by the 
presence of the nuclear medium. While this assumption is strongly supported by electron-nucleus scattering data in the quasi elastic channel, analyses of neutrino DIS data are often 
carried out allowing for a medium modification of the nucleon structure functions \cite{petti,haider}, or of the parton distributions entering their definitions \cite{kumano}.

The approach of Refs. \cite{petti,haider} makes use of a model of the nuclear spectral function, and includes a variety of medium effects, such as the 
$\pi$- and $\rho$- meson cloud contributions and nuclear shadowing. The authors of Ref.~\cite{kumano}, on the other hand, provide a parametrization of the nuclear 
parton distributions at order $\alpha_s$ obtained from a fit to the measured nuclear cross sections.

\section{Interpretation of the CCQE cross section}
\label{interpretation}

The data set of CCQE events collected by the MiniBooNE collaboration \cite{BooNECCQE} provides an unprecedented opportunity to carry out a systematic
study of the double differential cross section of the process,
\beq
\nu_\mu + ^{12}\mkern -5mu C \rightarrow \mu^- + X \ ,
\eeq
averaged over the neutrino flux shown in the right panel of Fig. \ref{fluxav}. 

As pointed out in the previous Section, the CCQE neutrino-nucleon process is described in terms of three
form factors. The proton and neutron electromagnetic form factors, which have been precisely measured
up to large values of $Q^2$ in electron-proton and electron-deuteron scattering experiments,
and the nucleon axial form factor $F_A$, parametrized in terms of the axial mass $M_A$ as in Eq. \eqref{FA}.
The data analysis performed using the RFGM yields an axial mass $M_A \approx 1.35$ GeV, significantly larger than that obtained from deuteron 
data \cite{bernard,bodek2}. A large value of the axial mass, $M_A \approx 1.2$ GeV, has been also reported by the analysis of the CCQE neutrino-oxygen 
cross section carried out by the K2K collaboration \cite{K2K}, while the NOMAD collaboration released the value $M_A = 1.05$ GeV, compatible with 
the world average of deuteron data, resulting from the analysis of CCQE neutrino- and antineutrino-carbon interactions at larger beam energies 
($E_\nu \sim 10$ GeV) \cite{NOMAD}.

\begin{figure}[h!]
\includegraphics[scale= 0.50]{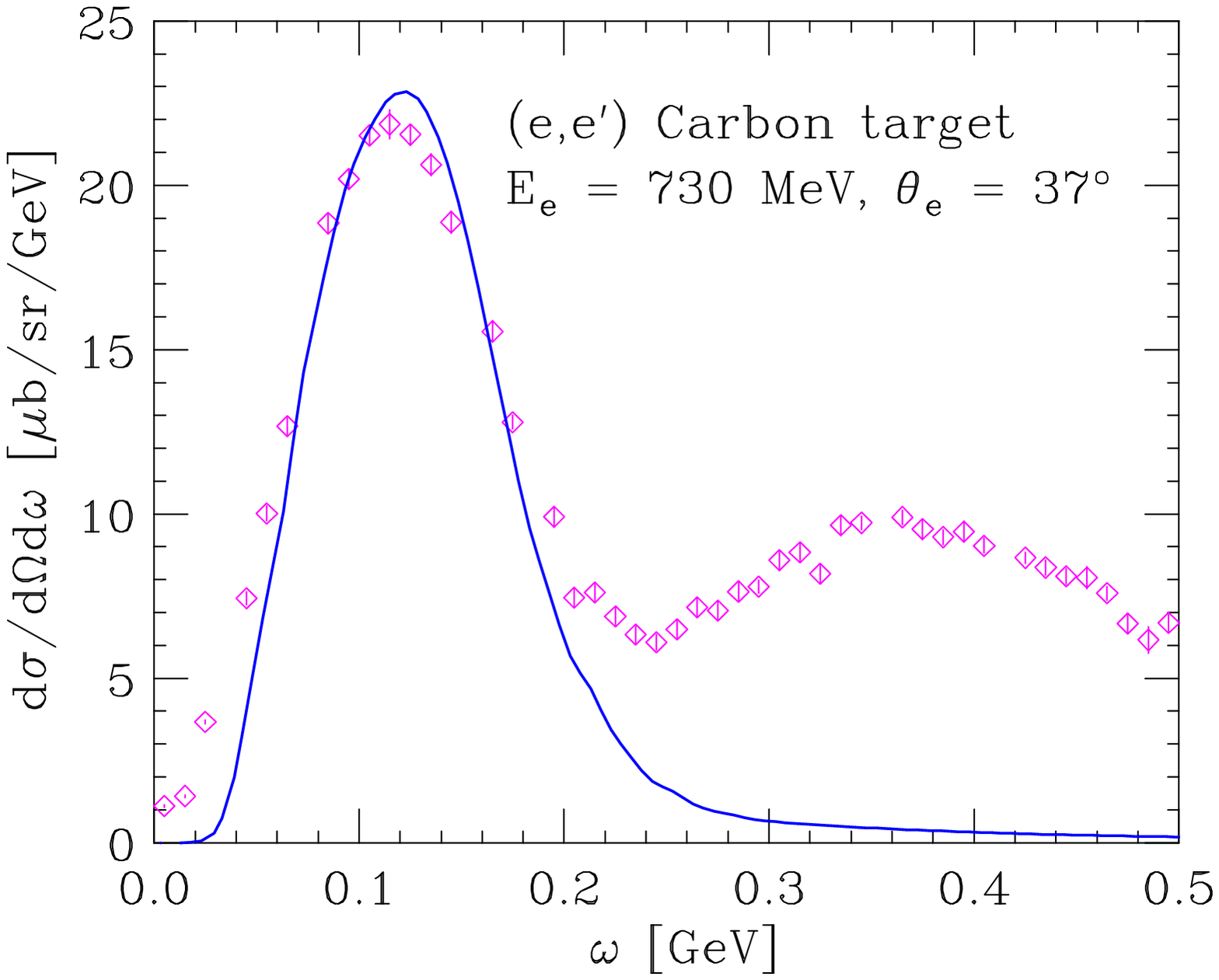} \hspace*{.5in}  \includegraphics[scale= 0.60]{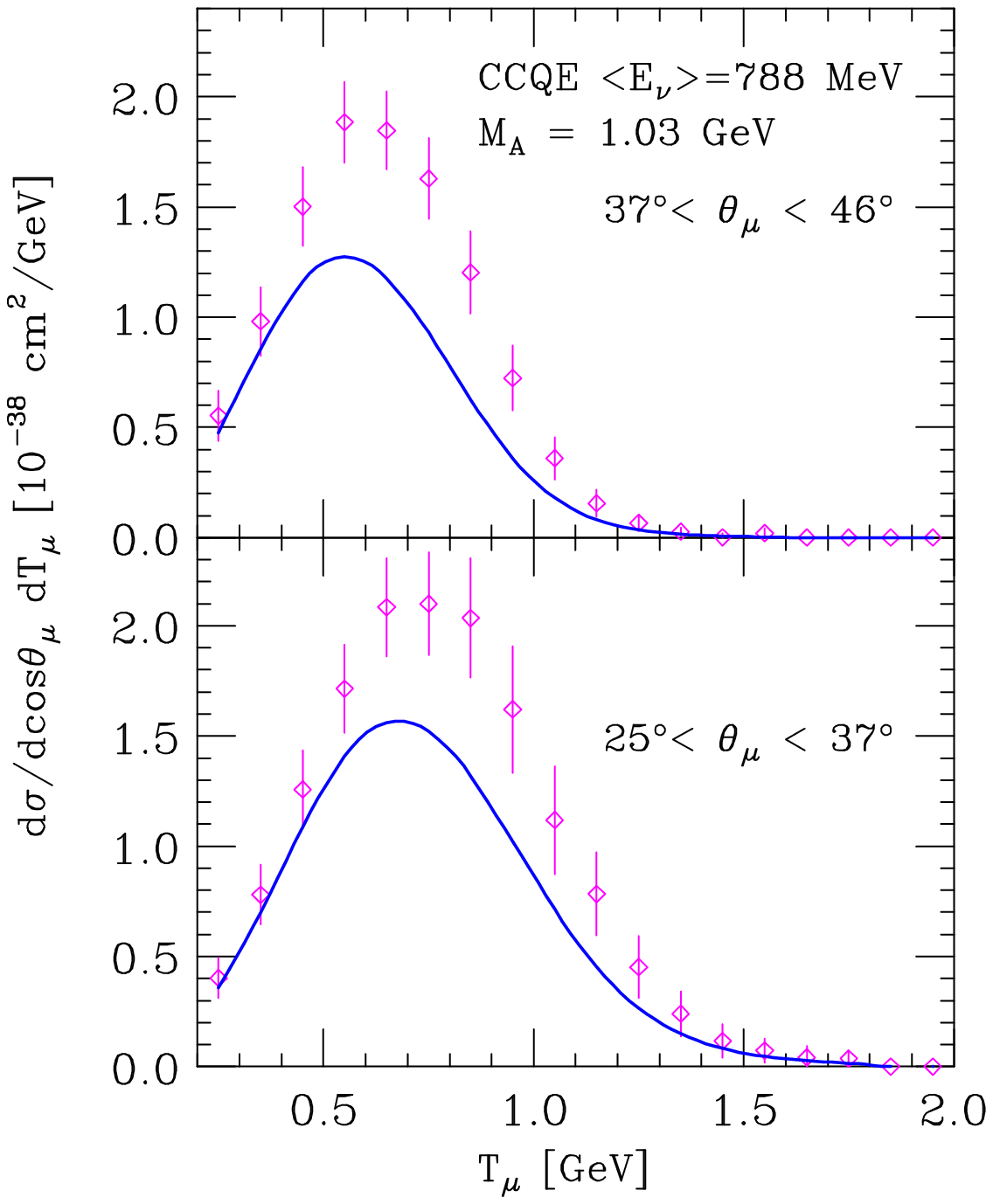} 
\vspace*{-.1in}
\caption{Left panel: inclusive electron-carbon cross section at beam energy $E_e=$ 730 MeV and electron scattering
angle $\theta_e=37^\circ$, plotted as a function of the energy loss $\omega$. The data points are taken from
Ref. \cite{12C_ee}.
 Right panel: Flux averaged double differential CCQE cross section measured by the MiniBooNE collaboration
\cite{BooNECCQE},  shown as a function of the kinetic energy of the outgoing muon. The upper and lower panels correspond to
to different values of the muon scattering angle. Theoretical results have been obtained using the same spectral
functions and  vector form factors employed in the calculation of the electron scattering cross section of the left panel,
and a dipole parametrizaition of the axial form factor with $M_A=1.03$ GeV.}
\label{compare}
\end{figure}

 It would be tempting  to interpret the value of $M_A$ reported by MiniBoonNE as an {\em effective} axial mass, modified by nuclear 
effects not included in the RFGM. However, theoretical studies carried out within the IA scheme with a realistic carbon spectral function 
-- an approach that has proved capable of providing a quantitative account of a wealth of electron scattering data in the quasi elastic sector --
fail to describe the flux averaged double differential cross section of Ref. \cite{BooNECCQE}. This striking feature is illustrated in Fig. \ref{compare}. 
The left panel shows a comparison between the electron scattering data data of Ref. \cite{12C_ee} and the results obtained using the spectral function of Ref. \cite{LDA}, 
while in the right panel the results obtained within the same scheme and setting $M_A=1.03$ MeV are compared to the flux averaged double differential CCQE cross 
section measured by the MiniBooNE collaboration,  shown as a function of kinetic energy of the outgoing muon \cite{coletti}.
It is apparent that height, position and width of the QE peak measured in electron scattering, mostly driven by the energy and momentum dependence of the
spectral function, are well reproduced, while the peaks exhibited by the neutrino cross sections are largely underestimated.

The authors of Ref. \cite{coletti} argued that the differences observed in Fig. \ref{compare} are to be largely ascribed to the flux average involved in the determination of the neutrino 
cross section, leading to the appearance of contributions of reaction mechanisms not taken into account in the IA picture. To overcome this difficulty, they advocated the development of
models based on a {\em new paradigm}, in which all relevant reaction mechanisms are {\em consistently} taken into account within a unified description of nuclear dynamics.
While in this review we will mainly focus on the approach based on the spectral function formalism, it has to be mentioned  that
a unified description of a variety of nuclear effects can be also obtained within a completely different framework, based on transport theory (for a recent review see, e.g., Ref~\cite{buss}).   

\subsection{Role of reaction mechanisms other than single nucleon knockout}

In MiniBooNE data analysis, an event is labeled as CCQE if no final state pions are detected in addition to the outgoing muon.
The simplest reaction mechanism compatible with this definition is single nucleon knockout, induced by the one-nucleon 
contributions to the nuclear current (see Eq. \eqref{nuclear:current}). In the absence of NN correlations the spectator 
$(A-1)$-particle system is left in a bound state, and the final nuclear state, consisting of the knocked out nucleon and the recoiling 
residual nucleus, is said to be a one particle-one hole state. 

It has been suggested that the observed excess of CCQE cross section may be traced back to the occurrence of events 
with two particle-two hole final states \cite{martini,nieves}. According to the above definition, these events 
cannot be distinguished from those with one particle-one hole final states . Therefore, they are often referred to as CCQE-like.
The role of two particle-two hole interactions at higher energies, up to 10 GeV, has also been recently discussed in Ref. \cite{Gran2}.

It has to be pointed out, however, that the approaches of Refs. \cite{martini,nieves}, while including two-nucleon current 
contributions, are based on the oversimplified independent particle model (IPM) of nuclear structure, the deficiencies of which 
have long been recognized \footnote{ In their
classic Nuclear Physics book, first published in 1952, Blatt and Weisskopf warn the reader that ``the limitation of any independent particle
model lies in its inability to encompass the correlation between the positions and spins of the various particles in the
system'' \cite{BW}.}. This issue is of paramount importance for the interpretation of CCQE-like
events, since within the IPM two particle-two hole states can only be excited by two-nucleon 
meson-exchange currents (MEC). On the other hand, in the presence of NN correlations final states with two nucleons in the continuum 
may be also produced through two additional mechanisms, which {\em do not} involve two-nucleon currents: 
i) initial state correlations (ISC), and ii) final state interactions (FSI). 

A fully consistent analysis of the role of two particle-two hole final state within a realistic model of nuclear structure 
obviously requires that all mechanisms leading to the appearance of these final states be included, using a quantum-mechanical 
approach properly taking into account interference between the transition amplitudes involving one- and two-nucleon currents.  
Within the approach of Refs. \cite{martini,nieves}, based on a model of nuclear dynamics in which correlations are 
not taken into account, these interference terms are generated by adding "ad hoc" contributions to the two-body current \cite{torino}.

Within the IA, ISC are taken into account using realistic spectral functions, which include the contribution of the continuum spectrum 
associated with unbound states of the residual nucleus. Their main effect is the 
appearance of a tail of the cross section, extending to large $T_\mu$, clearly visible in the left panel of of Fig. \ref{compare}. 
Semi-inclusive $(e,e^\prime p)$ data \cite{daniela} suggest that this contribution is not large, amounting to $\sim 10$\% of the integrated spectrum. 
In principle, this reaction mechanism might be clearly identified detecting two nucleons moving in opposite directions with momenta much larger 
than the Fermi momentum $p_F \sim$ 250 MeV (see, e.g., Ref. \cite{subedi}).

In inclusive processes, FSI lead to a shift of the energy loss spectrum, arising from interactions between the knocked out nucleon and 
the mean field of the recoiling nucleus, and a redistribution of the strength from the quasi free bump to the tails, resulting  
from rescattering processes. Theoretical studies of electron-nucleus scattering suggest that in the kinematical region relevant to the MiniBooNE analysis
the former mechanism, which does not involve the appearance of two particle-two hole final states, dominates 
A recent discussion of the inclusion of FSI within the IA scheme can be found in Ref. \cite{benharFSI}). 
A different approach, based on a Monte Carlo simulation, is described in Ref. \cite{nieves_FSI}.

As advocated in Refs. \cite{martini,nieves}, the most important contribution involving two particle-two hole final states is likely to arise 
from processes involving MEC, the inclusion of which is long known to be needed to explain the measured nuclear electromagnetic response in the 
transverse channel \cite{schiavilla}.  

\begin{figure}[h!]
\includegraphics[scale= 0.6]{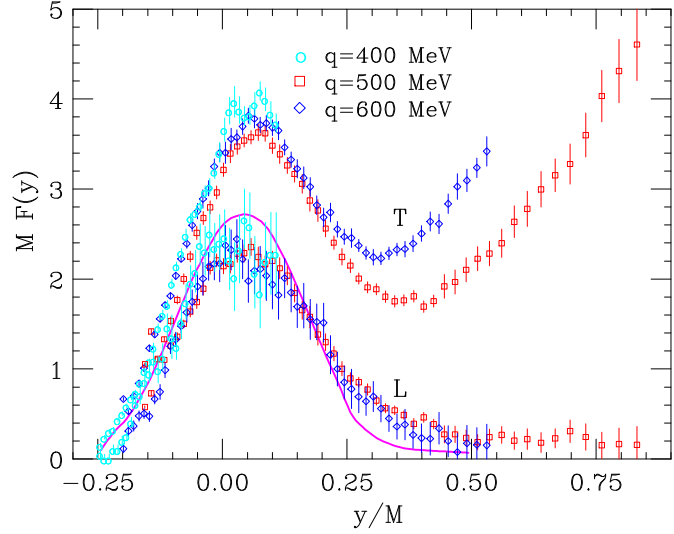} 
\vspace*{-.1in}
\caption{Longitudinal (L) and transverse (T) scaling functions of Carbon at $|{\bf q}| =$ 400, 500, and 600 MeV, resulting from the analysis of Ref. \cite{finn}, 
based on the data of Ref. \cite{barreau}. For comparison, the solid line shows the scaling function obtained from the IA using the spectral function of Ref. \cite{LDA}.} 
\label{scaling}
\end{figure}

The role of the two nucleon current in electron scattering is best illustrated by comparing the longitudinal and 
transverse $y$-scaling functions, shown in Fig. \ref{scaling}. Scaling in the variable $y$ follows from the dominance of one-nucleon processes, allowing 
one to write the equation expressing conservation of energy in a very simple form. As a consequence, in the limit of large momentum transfer the 
nuclear response, which is in general a function of both ${\bf q}$ and $\omega$, becomes a function of a single variable 
$y=y({\bf q}$, $\omega)$ \cite{scaling1,scaling2}. The occurrence of scaling provides a strong handle on the reaction mechanism, while
the observation of scaling violations reveals the role played by processes beyond the IA.  

Figure \ref{scaling} shows the scaling functions associated with the longitudinal (L) and transverse (T) responses of Carbon, extracted from 
electron scattering data \cite{finn,barreau}. The onset of scaling is clearly visible in the region of the quasi free peak, corresponding to $y \sim 0$, 
where the data points at different momentum transfer tend to sit on top of one another as $|{\bf q}|$ increases. On the other hand, large scaling violations, 
arising mainly from resonance production, appear in the transverse channel at $y>0$, corresponding to $\omega > Q^2/2M$. 
In addition, in the scaling region the transverse function turns out to be significantly 
larger than the longitudinal one, while within the IA picture the two scaling functions are expected to be identical. 

The results of highly accurate calculations carried out for light nuclei in the non relativistic regime strongly suggest that in the quasi elastic region 
single nucleon knockout processes are dominant in the longitudinal channel, while both one- and two-nucleon
mechanisms provide comparable contributions in the transverse channel \cite{CJSS}.

\begin{figure}[h!]
\includegraphics[scale= 0.50]{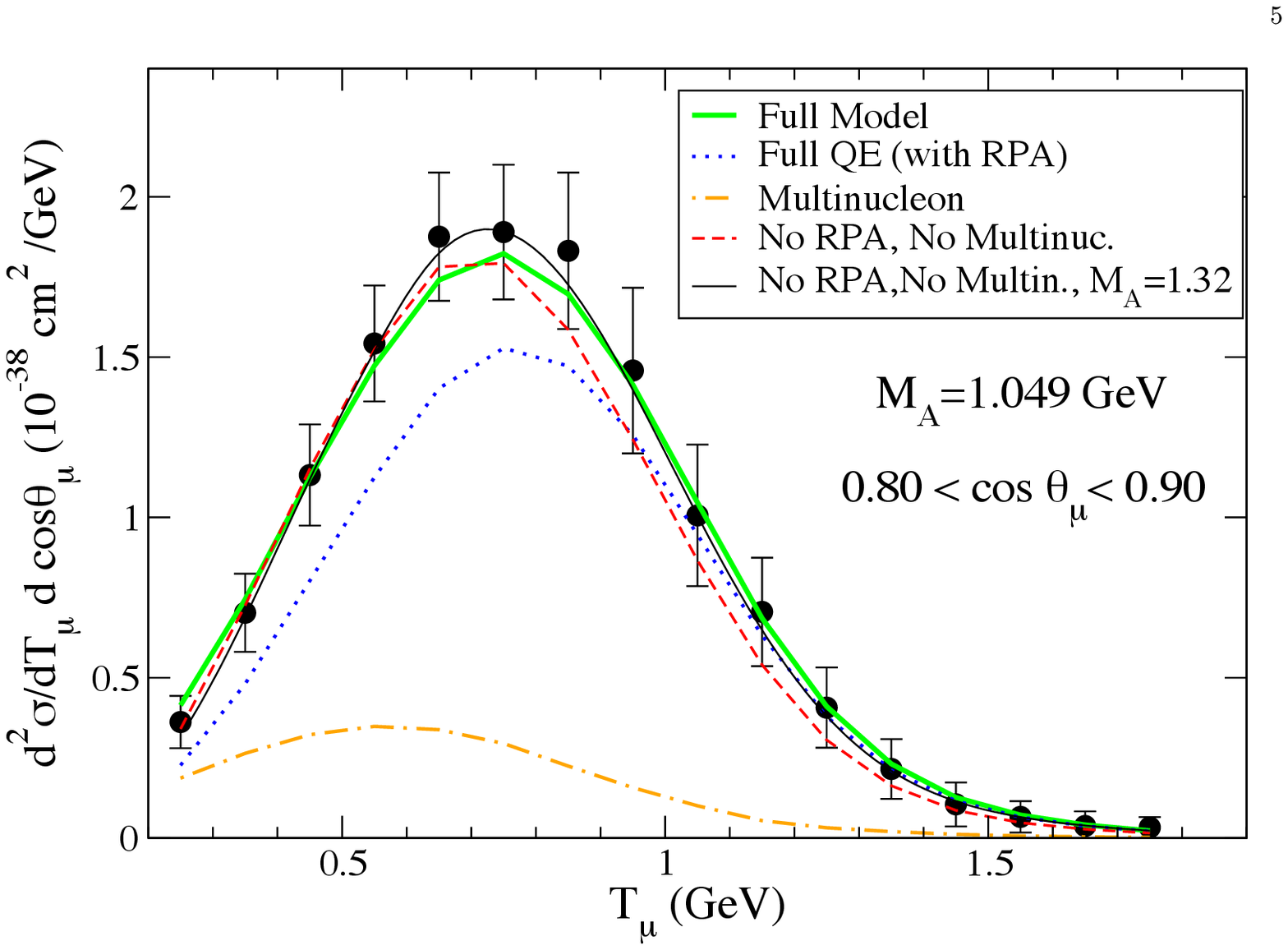} \hspace*{.2in}  \includegraphics[scale= 0.925]{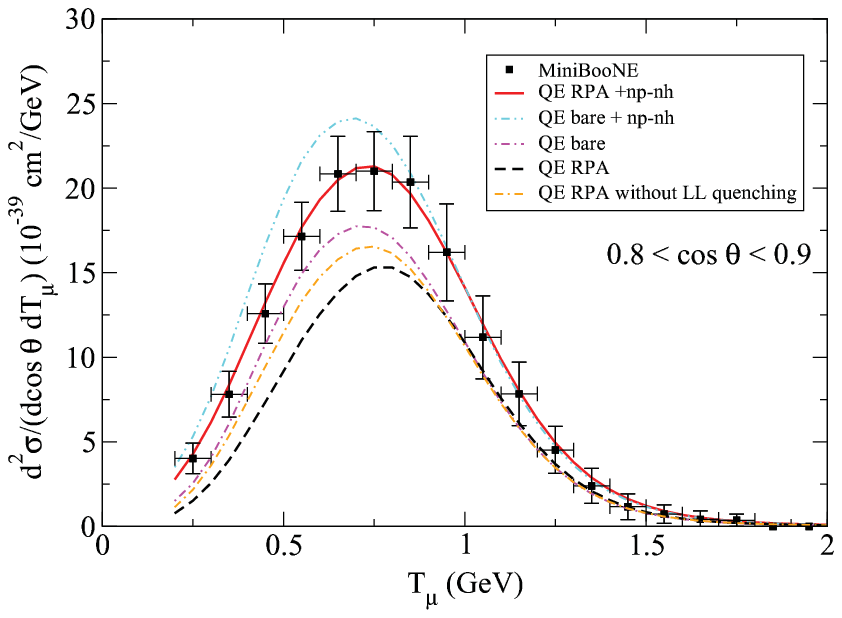} 
\vspace*{-.1in}
\caption{Comparison between the flux averaged muon energy spectrum 
at muon scattering angle $\theta_\mu$ such that $0.8 \leq \cos \theta_\mu  \leq 0.9$, measured by the MiniBooNE collaboration \cite{BooNECCQE} and the theoretical 
results of Refs. \cite{martini} (thick solid line of the right panel) and \cite{nieves} (thick solid line of the left panel). All theoretical calculations 
have been carried out using a value of the axial mass consistent with the one extracted from deuteron data.}
\label{numec1}
\end{figure}

The authors of Refs. \cite{martini,nieves} carried out extensive calculations of the CCQE neutrino-carbon cross section, averaged over the MiniBooNE flux, 
taking into account the effects of MEC as well as collective nuclear excitations, which are known to be important at 
low momentum transfer.  As an example, in Fig. \ref{numec1} the results of these approaches, obtained using a value of the axial mass consistent with 
the one extracted from deuteron data, are compared to the MiniBooNE muon energy spectrum 
at muon scattering angle $\theta_\mu$ such that $0.8 \leq \cos \theta_\mu \leq 0.9$. After inclusion of MEC, both schemes turn out to 
provide a quantitative account of the data, and the same pattern is observed for all values of $\theta_\mu$.

\begin{figure}[h!]
\includegraphics[scale= 0.65]{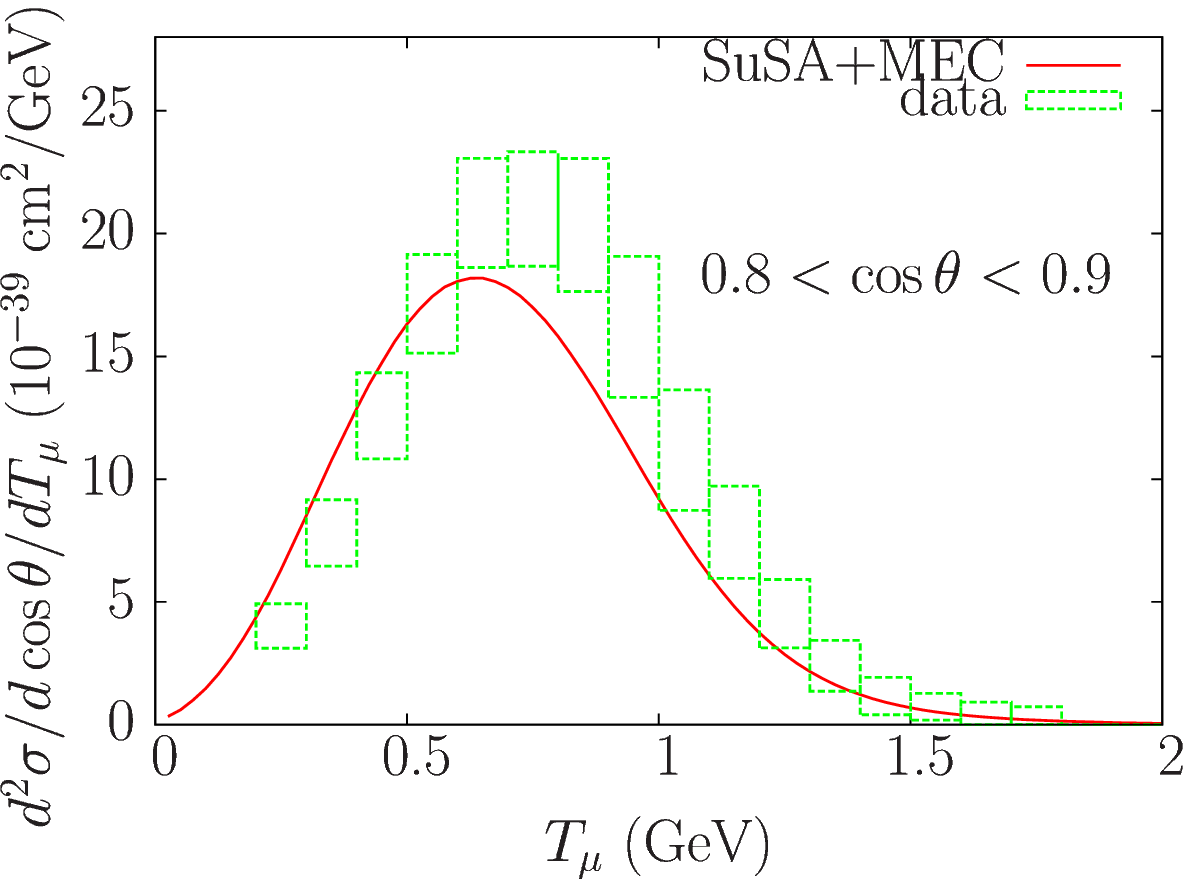} \hspace*{.3in}  \includegraphics[scale= 0.65]{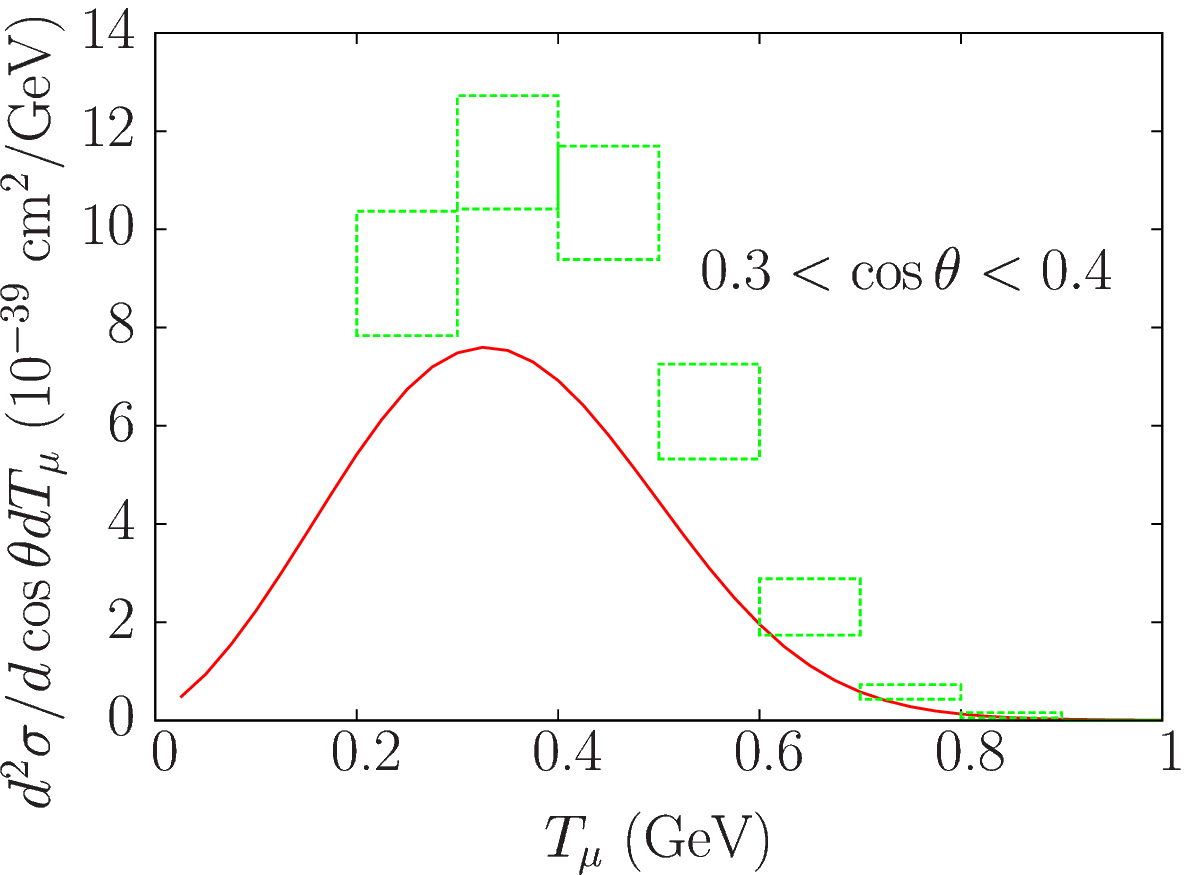} 
\vspace*{-.1in}
\caption{Comparison between the flux averaged muon energy spectra measured by the MiniBooNE collaboration
at muon scattering angle $\theta_\mu$ such that $0.8 \leq \cos \theta_\mu \leq 0.9$ (left panel) and  $0.3 \leq \cos \theta_\mu  \leq 0.4$ (right panel) \cite{BooNECCQE}
and the results of the approach of Ref. \cite{barbaro}. The data,  with their error bars, are represented by the boxes, while the solid lines show the
calculated spectra.}
\label{numec2}
\end{figure}

Figure \ref{numec2} shows a comparison between MiniBooNE data and the results of a different theoretical approach \cite{barbaro}. 
The authors of Ref. \cite{barbaro} developed a phenomenological procedure based on an extension of the $y$-scaling analysis, suitable 
to take into account the effects of processes involving MEC. It is apparent that, while at $0.8 \leq \cos \theta_\mu  \leq 0.9$ (left panel) 
inclusion of two-nucleon effects bring theory and experiment into agreement, at the larger angles corresponding to 
$0.3 \leq \cos \theta_\mu  \leq 0.4$ the measured cross section is still severely underestimated.

As pointed out above, a fully consistent treatment of processes involving two particle-two hole final states  requires a realistic model of
nuclear structure, taking into account the effects of NN correlations. Models including MEC within the framework of the IPM, 
such as those of Refs. \cite{martini,nieves}, are in fact based on the strong assumption that meson exchange, while playing an important role 
when the associated current is involved in interactions with an external probe, can be safely ignored in the description of the nuclear initial 
and final states.  The resulting description of the nuclear scattering process appears to be conceptually  inconsistent, although the impact of 
this issue on the numerical results needs to be carefully investigated. 

Going beyond this scheme in the kinematical region in which non relativistic approximations are not applicable requires an extension 
of the {\em factorization paradigm} underlying the IA, expressed by Eq.~\eqref{factorization}.

The starting point is the generalization of the ansatz of Eq. \eqref{factorization} for the hadronic final state to the case in which the interaction with the probe 
involves two-nucleons:
\begin{align}
|X\rangle \longrightarrow |{\bf p} \ {\bf p}^\prime\rangle \otimes |n_{(A-2)} \rangle  = | n_{(A-2)}; {\bf p} \ {\bf p}^\prime \rangle  \ , 
\end{align} 
where $ |n_{(A-2)} \rangle$ is the state of the spectator $(A-2)$-nucleon system, carrying momentum ${\bf p}_n$.

It follows that the matrix element of the two nucleon current simplifies to (compare to Eq. \eqref{fact:matel})
\begin{align}
\langle X | {j_{ij}}^{\mu} | 0 \rangle  \rightarrow 
 \int d^3k d^3 k^\prime M_n({\bf k},{\bf k}^\prime) \ 
\langle {\bf p} {\bf p}^\prime | {j_{ij}}^\mu | {\bf k} {\bf k}^\prime \rangle 
 \ \delta({\bf k} + {\bf k}^\prime - {\bf p}_n) \ ,
\end{align}
with the amplitude $M_n({\bf k},{\bf k}^\prime)$ given by
\begin{align}
\label{def:Mn}
M_n({\bf k},{\bf k}^\prime) = \langle n_{(A-2)}; {\bf k}  \  {\bf k}^\prime | 0 \rangle \ .
\end{align}
Within this scheme, the nuclear amplitude $M_n({\bf k},{\bf k}^\prime)$ turns out to be independent of ${\bf q}$, and can therefore be 
obtained within non relativistic many body theory without any problems. 

The connection with the spectral function formalism discussed in Section \ref{IA} becomes apparent noting that the two-nucleon spectral function 
$P({\bf k},{\bf k}^\prime,E)$, 
yielding the probability of removing {\em two nucleons} from the nuclear ground state leaving the residual system with excitation energy $E$, is
defined as 
\begin{align}
\label{def:pke2}
P({\bf k},{\bf k}^\prime,E) = \sum_n |M_n({\bf k},{\bf k}^\prime)|^2 \delta(E + E_0 - E_n) \ ,
\end{align}
where $M_n({\bf k},{\bf k}^\prime)$ is defined as in Eq. \eqref{def:Mn} and $E_0$ is the ground state energy. 

\begin{figure}[h!]
\includegraphics[scale= 0.5]{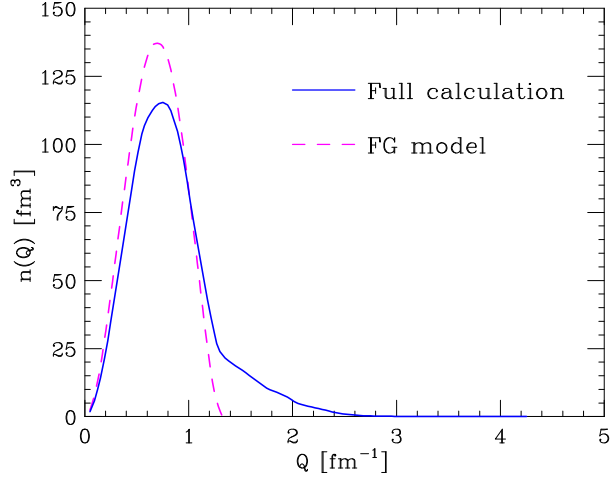} 
\vspace*{-.1in}
\caption{Comparison between the two-nucleon relative momentum distribution of nuclear matter computed within nuclear many-body theory using a realistic 
hamiltonian (solid line) \cite{spec2}, and the prediction of the Fermi gas (FG) model (dashed line).}
\label{reldist}
\end{figure}

The two-nucleon spectral function of uniform and isospin symmetric nuclear matter at equilibrium density has been calculated within 
nuclear many-body theory using a realistic hamiltonian \cite{spec2}. As an example, the resulting relative momentum distribution, defined as
\begin{align}
\label{rel:dist}
n({\bf Q}) = 4 \pi |{\bf Q}|^2 \int d^3 K \ n\left( \frac{ {\bf Q} }{2} + {\bf K}, \frac{ {\bf Q} }{2} - {\bf K} \right)
\end{align}
with
\begin{align}
n({\bf k},{\bf k}^\prime) = \int dE  \ P({\bf k},{\bf k}^\prime,E) \ , 
\end{align}
 and
 \begin{align}
 {\bf K} = {\bf k}+{\bf k}^\prime  \ \ \ , \ \ \ {\bf Q} = \frac{ {\bf k}-{\bf k}^\prime }{2} \ .
 \end{align}
 is shown by the solid line of Fig. \ref{reldist}. Comparison with the prediction of the Fermi gas model, represented by the dashed line, 
 indicates that correlation effects are sizable. As expected, they lead to a quenching of the peak of the distributions and an enhancement 
 of the high momentum tail.
 
Using the $(A-1)$- and $(A-2)$-nucleon amplitudes of Refs. \cite{BFF} and \cite{spec2}, ISC and MEC contributions to the nuclear cross sections can 
be calculated in a fully consistent fashion, taking into account interference and using the the fully relativistic expression of two-nucleon current. 
As pointed out above, the contribution of FSI is not expected to be 
critical in the kinematical region spanned  by MiniBooNE data. However, in principle they can be also taken into account within the spectral function 
formalism \cite{benharFSI}. 

\section{Neutrino energy reconstruction}
\label{reconstruction}

In recent years, nuclear effects in neutrino interactions, while being interesting in their own right, have been mainly studied to appraise
their impact on the determination of neutrino oscillations. As an example, in this Section we will discuss the
uncertainty on neutrino energy reconstruction arising from the the nuclear models 
employed in data analysis.

Let us consider, for simplicity, two-flavor mixing. The expression of the probability that a neutrino oscillate 
from flavor $\alpha$ to flavor $\beta$ after travelling a distance $L$ 
\begin{align}
P_{\nu_\alpha \to \nu_\beta} = \sin^2 2\theta \ \sin^2 \left( \frac{\Delta m^2 L}{4 E_\nu} \right) \ ,
\end{align}
where $\theta$ and $\Delta m^2$ are the mixing angle and the squared mass difference, respectively, clearly shows that 
the accurate determination of the neutrino energy, $E_\nu$, plays a critical role. A wrongly reconstructed $E_\nu$ does in fact result
in an incorrect determination of the mixing angle.

The starting point for neutrino energy reconstruction in $\nu_\mu$ CCQE interactions is the equation expressing the requirement that the 
scattering process be elastic, i.e.
\beq
\label{rec:1}
(k_\nu + p_n -k_\mu)^2= M_p^2 ,
\eeq
where $k_\nu \equiv(E_\nu, {\bf k}_\nu)$ and $k_\mu \equiv(E_\mu,{\bf k}_\mu)$ are the four momenta of the incoming neutrino and outgoing muon, 
respectively, $M_p$ is the proton mass and $p_n \equiv(E_n,{\bf p}_n)$, with $E_n = M_A - E_{A-1}$ and 
$E_{A-1} = \sqrt{ (M_A - M_n + E)^2 + |{\bf p}_n|^2}$, 
is the four momentum struck neutron.

From Eq. \eqref{rec:1} it follows that 
\beq
\label{rec:2}
E_\nu= \frac{M_p^2 - m_\mu^2 -E_n^2 +2E_\mu E_n -2\m{k}_\mu\cdot\m{p}_n+|\m{p}_n|^2}{2(E_n-E_\mu +|\m{k}_\mu| \cos\theta_\mu-|\m{p}_n| \cos\theta_n)} \ ,
\eeq
where $\theta_\mu$ is the muon angle relative to the neutrino beam and $\cos\theta_n= (\m{k}_\nu\cdot\m{p}_n)/(|\m{k}_\nu||\m{p}_n|)$.
The  above equation clearly shows that $E_\nu$ is not uniquely determined by the measured kinematical variables, $E_\mu$ and $\theta_\mu$, 
but exhibits a distribution reflecting the energy and momentum distribution of the struck neutron. Therefore, it depends on the nuclear model employed to 
describe the target ground state.

In the analysis of MiniBooNE data \cite{BOONE}, the energy of the incoming neutrino has been reconstructed by setting $|\m{p}_n|=0$ and 
fixing the neutron removal energy to a constant value, i.e. setting $E=\epsilon$, implying in turn 
$E_n= M_n - \epsilon$. The resulting expression is
\beq
\label{ereconstructed}
E_\nu^{\rm rec}= \frac{2(M_n-\epsilon)E_\mu -(\epsilon^2 - 2M_n\epsilon +m_\mu^2 + \Delta M^2) }{2[M_n - \epsilon -E_\mu + |\m{k}_\mu| \cos\theta_\mu]}
\eeq 
where $M_n$ is the neutron mass, $\Delta M^2= M_n^2 -M_p^2$, and $|\m{k}_\mu|=\sqrt{E_\mu^2 - m_\mu^2}$
 is the magnitude of the three-momentum of the outgoing muon.
 
In general, the neutrino energy distribution, $F(E_\nu)$, can be obtained from Eq.(\ref{rec:2}) using 
values of $|{\bf p}_n|$ and $E$ sampled from the probability distribution
$|{\bf p}_n|^2 P({\bf p}_n,E)$, and assuming that the polar and azimuthal angles specifying
the direction of the neutron momentum be uniformly distributed.

\begin{figure}[h!]
\includegraphics[scale= 0.65]{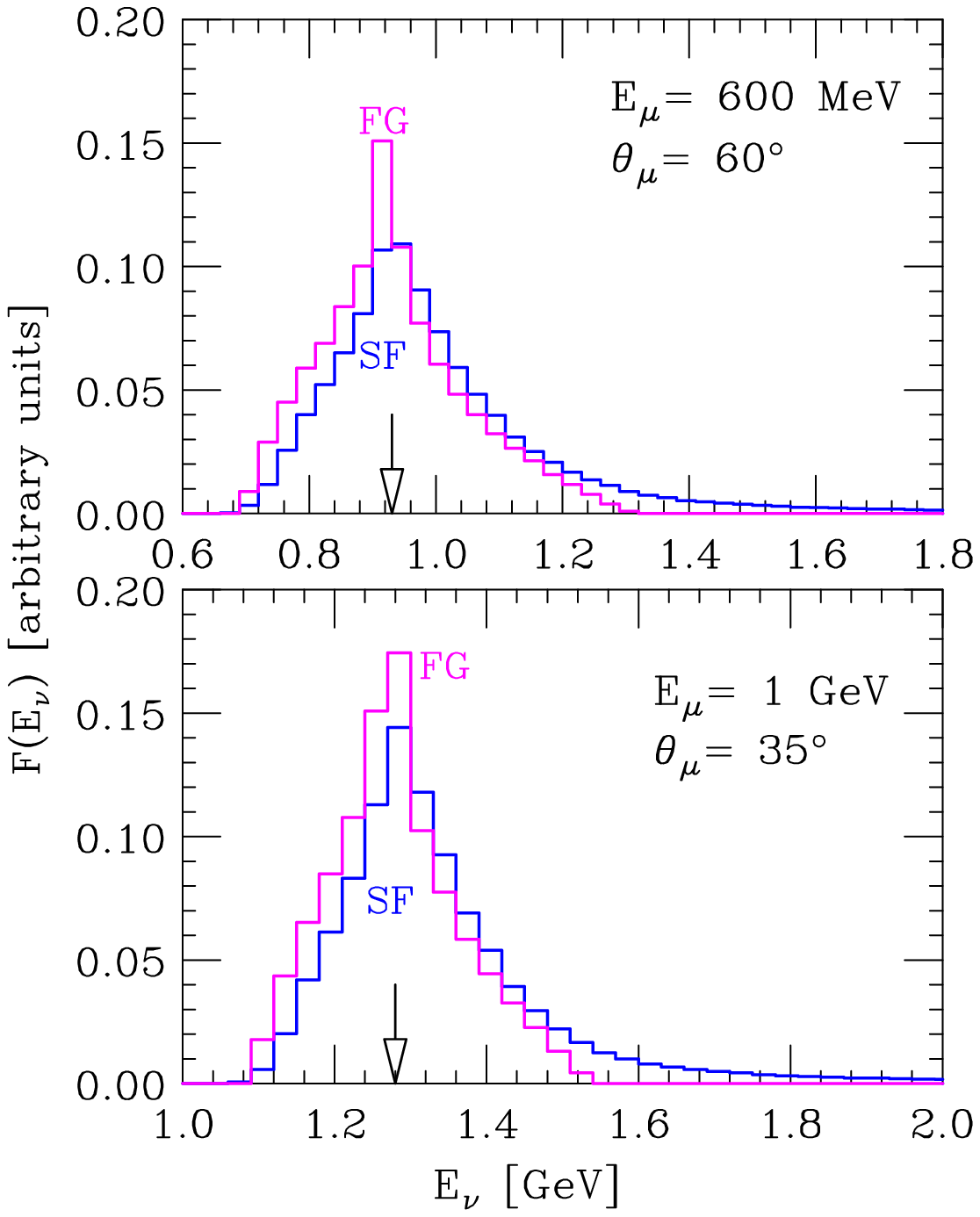} \hspace*{.3in}  \includegraphics[scale= 0.65]{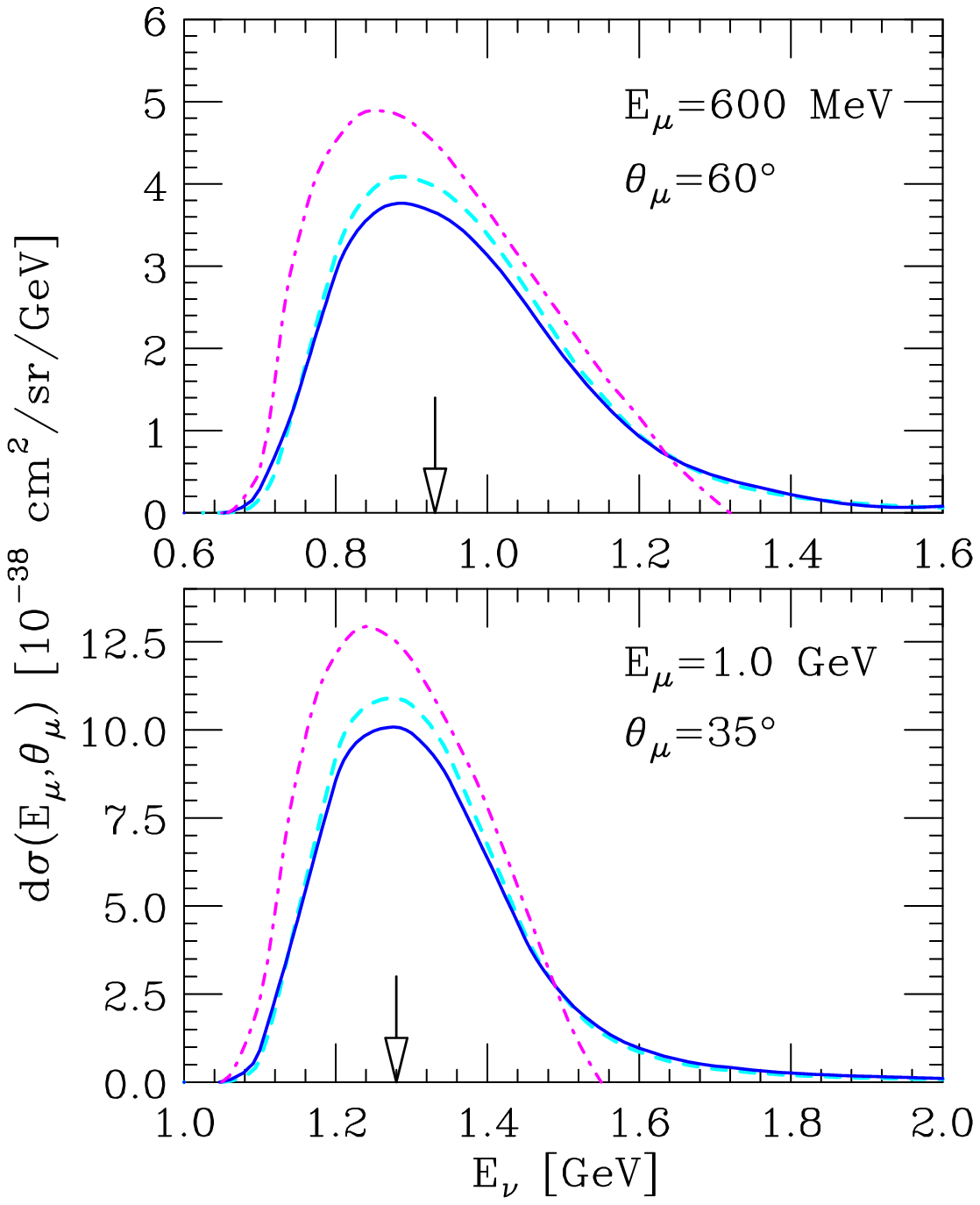} 
\vspace*{-.1in}
\caption{Left: neutrino energy distribution at $E_\mu=$ 600 MeV and $\theta_\mu=$ 60 deg (upper panel) and 
$E_\mu=$ 1 GeV and $\theta_\mu=$ 35 deg (lower panel), reconstructed from Eq.(\ref{rec:2})
using 2 $\times$10$^4$ pairs of ($|{\bf p}_n|,E$) values sampled from the probability distributions associated with the oxygen spectral function
of Ref.\cite{spec2} (SF) and the RFGM, with Fermi momentum $p_F=$ 225
MeV and removal energy $\epsilon=27$ MeV (FG). The arrows point to
the values of $E_\nu^{\rm rec}$ obtained from Eq. (\ref{ereconstructed}). Right: differential cross section of the process
$\nu_\mu + A \rightarrow \mu + p + (A-1)$, at $E_\mu=$ 600 MeV and $\theta_\mu=$ 60 deg (upper panel) and $E_\mu=$ 1 GeV and
$\theta_\mu=$ 35 deg (lower panel), as a function of the incoming neutrino energy.
 The solid line shows the results of the
full calculation, carried out within the approach of Refs. \cite{gangofsix,LDA},
whereas the dashed line has been obtained neglecting the effects
of FSI. The dot-dash line corresponds to the
 RFG model with Fermi momentum $p_F=$ 225 MeV and
removal energy $\epsilon=27$ MeV. The arrow points to
the value of $E_\nu^{\rm rec}$ obtained from Eq. (\ref{ereconstructed}). }
\label{sampling}
\end{figure}

To gauge the effect of the high momentum and high removal energy tails 
of the spectral functions obtained within realistic dynamical approaches including NN correlations, 
the authors of Ref. \cite{axmass} have compared the $F(E_\nu)$ computed
using 2 $\times$10$^4$ pairs of ($|{\bf p}_n|,E$) values drawn from the probability
distributions associated with the oxygen spectral function of Ref.\cite{PRD}, to that 
obtained from the RFGM, with Fermi momentum $p_F=$ 225 MeV and removal energy $\epsilon=27$ MeV.
The results corresponding to $E_\mu=$~600~MeV and $\theta_\mu=$~35 deg, and 
$E_\mu=$~1~GeV and $\theta_\mu=$~35 deg, are displayed in the left
panels of Fig.~\ref{sampling}.

The distributions predicted by the RFGM model are more sharply
peaked at the neutrino energy given by Eq.(\ref{ereconstructed}), while the
$F(E_\nu)$ obtained from the spectral function of Ref.\cite{LDA}
are shifted towards higher energy by $\sim$ 20 MeV, with respect to the RFG results, and
exhibit a tail extending to very large values of $E_\nu$.

Note that the histograms of Fig. \ref{sampling} have been obtained from
Eq.(\ref{rec:2}), which in turn follows from the requirement of quasi elastic kinematics,
Eq.(\ref{rec:1}). However, the reconstruction of the neutrino energy from the measured
muon energy and scattering angle is also affected by the occurrence of FSI between the outgoing 
proton and the spectator nucleons.

In order to assess the total impact of replacing the RFGM with the approach
of Ref.\cite{gangofsix,LDA}, including both ISC and FSI,  the authors of Ref. \cite{axmass} 
have also computed the differential cross section
of the process $\nu_\mu + A \rightarrow \mu + p + (A-1)$, as a function of the
incoming neutrino energy $E_\nu$, for the muon kinematics of the left panel of Fig. \ref{sampling}.
The solid lines in the right panels of Fig.~\ref{sampling} show the results of the full calculation, carried out
using the spectral function of Ref.\cite{LDA}, while the  dashed lines have been obtained
neglecting the effects of FSI and the dot-dash lines correspond to the RFGM. Note that,
unlike the histograms of the left panels, the curves displayed in the right panels of 
Fig. \ref{sampling} have {\rm different} normalizations. 

 The differences between the results of the approach obtained from nuclear many-body theory and those
of the RFG model turn out to be sizable. The overall shift towards high energies and the
tails at large $E_\nu$ appearing in the histograms of the left panels Fig. \ref{sampling}, are still clearly
visible and comparable in size in the cross section, while the quenching with respect to the RFGM is larger. 

The reconstruction of neutrino energy in CCQE-like processes is more complex, as the four-momentum transfer is shared 
between two nucleons. As a consequence, it requires the knowledge of the two-nucleon spectral function of Eqs. \eqref{def:Mn} and \eqref{def:pke2}, and
the inclusion of correlation effects is essential. 

\begin{figure}[h!]
\includegraphics[scale= 0.625]{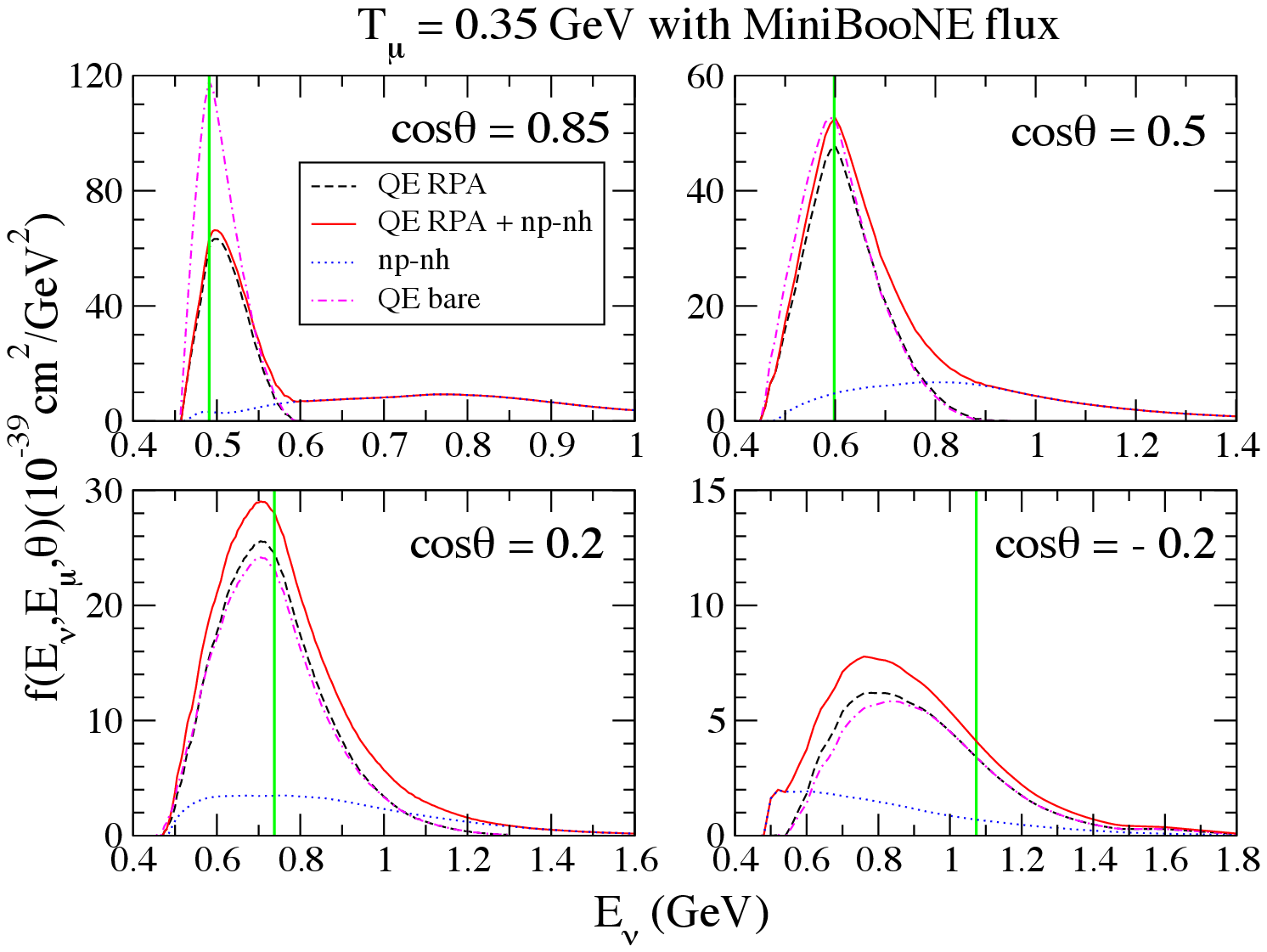} \hspace*{.0in}  \includegraphics[scale= 0.90]{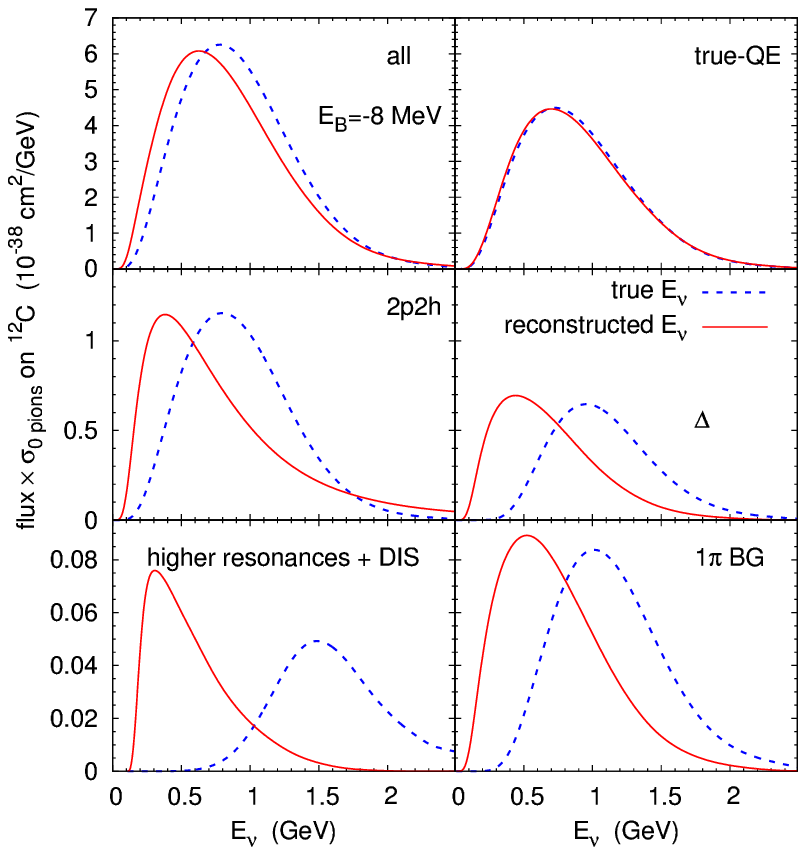} 
\vspace*{-.1in}
\caption{Left: Neutrino energy distributions at  $T_\mu =$ 350 MeV,  computed at different muon scattering angles 
using the MiniBooNE flux \cite{enu1}. The vertical lines correspond to the reconstructed $E_\nu$ of Eq. \eqref{ereconstructed}. Right: event distribution of zero 
pion events in the MiniBooNE experiment. The dashed curves show the distributions $\Phi(E_\nu)\sigma_{0\pi}(E_\nu)$, defined as in Eq. \eqref{enu:mosel}, 
corresponding to different reaction mechanisms, 
while the distributions $\Phi(E_\nu^{\rm rec}){\widetilde \sigma}_{0\pi}(E_\nu^{\rm rec})$ are displayed as solid lines \cite{enu2}.}
\label{fig:enudist}
\end{figure}
 
The uncertainty associated with the reconstruction of $E_\nu$ has been recently estimated in Ref. \cite{enu1} taking into account all reaction mechanisms 
included in the model of Ref. \cite{martini}.  The authors of Ref. \cite{enu1} computed the neutrino energy distribution, defined by the equation 
\begin{align}
\label{enu:martini}
f(E_\nu, E_\mu, \cos \theta_\mu) d E_\nu = C \left( \frac{ d \sigma }{ dE_\mu d\cos \theta_\mu } \right) \Phi(E_\nu) d E_\nu \ , 
\end{align}
where $\Phi(E_\nu)$ denotes the flux of incoming neutrinos with energy $E_\nu$ and the normalization constant $C$ is defined through
\begin{align}
C^{-1} = \int dE_\nu  \Phi(E_\nu) \ .
\end{align}
The results obtained using the MiniBooNE flux and setting $T_\mu = 350$ MeV are shown in the left panels of Fig. \ref{fig:enudist} for different values of the muon scattering 
angle. Comparison between the dot-dash line, representing the result obtained taking into account single nucleon knock out processes only, 
and the solid line, corresponding to the full calculation, shows that two-nucleon mechanism are important, and their effect on the neutrino energy 
reconstruction exhibits a strong dependence on $\cos \theta_\mu$.

The impact of reaction mechanisms other than single nucleon knock out on neutrino energy reconstruction has been also analyzed  in Refs. \cite{enu2,enu3}.
The authors of Ref. \cite{enu2} pointed out that the measured cross section extracted from events with no pions in the final state -- identified 
as CCQE in both the MiniBooNE and K2K analyses -- is obtained by dividing the event distribution at a given reconstructed energy by the incoming
neutrino flux at the same energy. This quantity, denoted ${\widetilde \sigma}_{0\pi}(E_\nu^{\rm rec})$,  is in general different from the {\em true} cross section 
evaluated at the {\em true} neutrino energy,   $\sigma_{0\pi}(E_\nu)$, which can be obtained from the relation 
\beq
\label{enu:mosel}
\Phi(E_\nu) \sigma_{0\pi}(E_\nu )= \int \mathcal{P}(E_\nu | E_\nu^{\rm rec})\Phi(E_\nu^{\rm rec}){\widetilde \sigma}_{0\pi}(E_\nu^{\rm rec}) \ .
\eeq
where $\mathcal{P}(E_\nu | E_\nu^{\rm rec})$ is the probability density of finding the energy $E_\nu$ in a distribution of events having the 
the same reconstructed energy $E_\nu^{\rm rec}$.
 
The right panels of Fig. \ref{fig:enudist} provide an illustration of the main results of Ref. \cite{enu2}. The dashed curves represent the distributions 
$\Phi(E_\nu)\sigma_{0\pi}(E_\nu)$, defined as in Eq. \eqref{enu:mosel}, while the distributions $\Phi(E_\nu^{\rm rec}){\widetilde \sigma}_{0\pi}(E_\nu^{\rm rec})$ 
are displayed as solid lines. Comparison between different reaction mechanisms shows that for processes involving the two nucleon current the difference between the solid 
and dashed lines is large. Sizable effects are are also visible in the resonance and pion production channels, although these mechanisms turn out to provide smaller 
contributions to the cross section.

\section{Summary and Conclusions}
\label{conclusions}

Over the past few years, the availability of the double-differential CCQE cross section measured by the MiniBooNE collaboration and the 
results of a new generation of theoretical studies have led to a better understanding of neutrino-nucleus interactions in a broad kinematical
range, as well as to the identification of a number of outstanding unresolved issues.

In view of the fact that no convincing evidence of medium modifications of the nucleon electromagnetic form factors has yet emerged, 
the excess of CCQE events in carbon reported by the MiniBooNE collaboration \cite{BooNECCQE}, the explanation of which 
within the RFGM requires a large increase of the nucleon axial mass with respect to the value obtained from deuteron data, 
is likely to be ascribable to the occurrence of processes other than single nucleon knock out, as advocated in Ref. \cite{coletti}.

The authors of Ref. \cite{martini,nieves} argued that the most important competing mechanism is
multinucleon knock out, leading to the appearance of two particle-two-hole final states that cannot be distinguished from 
the one particle-one hole final states associated with single nucleon knockout.

The models developed in Refs. \cite{martini,nieves}, while proving quite successful in explaining the MiniBooNE neutrino data
in terms of processes involving the two-nucleon current, are based on a somewhat oversimplified description of the nuclear initial and 
final states, in which the effects of correlations are not taken into account. One important consequence of this treatment 
of nuclear dynamics is that reaction mechanisms triggered by the one-nucleon current and producing two particle-two-hole final states
are not taken into account. However, these mechanisms and those involving the two-nucleon current are inextricably related to one another
and give rise to interference contributions to the cross section. Therefore,   they should be all consistently included. 

The main problem associated with a fully consistent description of nuclear structure and dynamics in the kinematical region relevant 
to neutrino experiments such as MiniBooNe and Miner$\nu$a stems from the large values of the momentum transfer, which make 
non relativistic approaches inapplicable. The extension of the factorization scheme underlying the spectral function formalism to the 
case of two-nucleon processes may provide a viable approach to overcome this difficulty and study interference effects neglected in the existing
calculations.

The dependence of the contribution of processes involving the two-nucleon current on the kinematical conditions should also be 
carefully investigated. It has been suggested that this analysis may help to shed light on the source of the large disagreement between the values of 
the nucleon axial mass reported by the MiniBooNE and NOMAD collaborations \cite{benhar_nufact11}.

The systematic study of the impact of nuclear effects on the determination of neutrino oscillation parameters is still in its infancy \cite{impact1,impact2}, 
and is likely to become a most active research field in the coming years. The problem of neutrino energy reconstruction, that plays a critical role in this 
context, has recently been analyzed using a variety of models including different reaction mechanisms. The emerging picture suggests that 
the reconstructed energy may turn out to be shifted towards lower values by as much as $\sim 100$ MeV,  with respect to the true energy. The authors of
Ref. \cite{enu2} argued that this uncertainty may hamper the extraction a CP violating phase from an oscillation result.

\begin{acknowledgements}
Some of the results discussed in this paper have been obtained in Collaboration with Artur M. Ankowski and Davide Meloni. 
The authors are also indebted to Camillo Mariani and Makoto Sakuda for many illuminating discussions. 
\end{acknowledgements}

\end{document}